\documentclass[conference]{IEEEtran}
\pdfoutput=1
\IEEEoverridecommandlockouts

\usepackage[cmex10]{amsmath}
\usepackage{cite}
\usepackage{xcolor}

\usepackage{amssymb}
\usepackage{amsfonts}
\usepackage{pifont}% http://ctan.org/pkg/pifont

\usepackage{enumitem}
\usepackage{path}
\usepackage{verbatim}
\usepackage{algorithm}
\usepackage{algorithmic}
\usepackage{framed}
\usepackage{footnote}
\usepackage{color}
\usepackage{amsmath}
\usepackage{cases}
\usepackage{subfigure}
\usepackage{theorem}
\usepackage{xurl}
\usepackage{acro}
\usepackage{eurosym}
\usepackage{multirow}

\usepackage[normalem]{ulem}
\usepackage{tikz}

\usepackage{tabularx}
\usepackage{booktabs}
\usepackage[colorlinks=true, allcolors=black]{hyperref}
\usepackage{threeparttable}

\usepackage{graphicx}

\usepackage{pgfplots}
\usepackage{pgfplotstable}
\pgfplotsset{compat=newest}
\graphicspath{{graphics/}}

{ \theorembodyfont{\normalfont} %\theorembodyfont{\rmfamily}

}

\newcounter{enumctr}

\DeclareFontFamily{U}{mathx}{\hyphenchar\font45}
\DeclareFontShape{U}{mathx}{m}{n}{<-> mathx10}{}
\DeclareSymbolFont{mathx}{U}{mathx}{m}{n}
\DeclareMathAccent{\widebar}{0}{mathx}{"73}

\newcolumntype{C}{>{\centering\arraybackslash}X} % centered version of "X" type
\DeclareAcronym{ann}{
    short = ANN,
    long  = Artificial Neural Networks,
}
\DeclareAcronym{api}{
    short = API,
    long  = Application Programming Interface,
}
\DeclareAcronym{arima}{
    short = ARIMA,
    long  = Auto-Regressive Integrated Moving Average,
}
\DeclareAcronym{alstm}{
    short = ALSTM,
    long  = Attention-based Long Short-Term Memory,
}
\DeclareAcronym{arnn}{
    short = ARNN,
    long  = Attention-based Recurrent Neural Networks,
}
\DeclareAcronym{astgcn}{
    short = ASTGCN,
    long  = Attention-based Spatio-Temporal Graph Convolutional Network,
}
\DeclareAcronym{cce}{
    short = CCE,
    long  = Categorical Cross Entropy,
}
\DeclareAcronym{cnn}{
    short = CNN,
    long  = Convolutional Neural Networks,
}
\DeclareAcronym{convlstm}{
    short = Conv-LSTM,
    long  = Convolutional Long Short-Term Memory,
}
\DeclareAcronym{dl}{
    short = DL,
    long  = Deep Learnining,
}
\DeclareAcronym{dpnst}{
    short = DPNst,
    long  = Destination Prediction Network based on Spatio-Temporal data
}
\DeclareAcronym{ebike}{
    short = e-bike,
    long  = Electric Bike,
}
\DeclareAcronym{em}{
    short = EM,
    long  = Electric Micromobility,
}
\DeclareAcronym{escooter}{
    short = e-scooter,
    long  = Electric Scooter,
}
\DeclareAcronym{esms}{
    short = ESMS,
    long  = Electric Shared Micromobility Services,
}
\DeclareAcronym{ev}{
    short = EV,
    long  = Electric Vehicle,
}
\DeclareAcronym{gcn}{
    short = GCN,
    long  = Graph Convolutional Neural Networks,
}
\DeclareAcronym{gdpr}{
    short = GDPR,
    long  = General Data Protection Regulation,
}
\DeclareAcronym{gtwr}{
    short = GTWR,
    long  = Geographically and Temporally Weighted Regression,
}
\DeclareAcronym{gps}{
    short = GPS,
    long  = Global Positioning System,
}
\DeclareAcronym{gru}{
    short = GRU,
    long  = Gated Recurrent Unit,
}
\DeclareAcronym{gwr}{
    short = GWR,
    long  = Geographically Weighted Regression,
}
\DeclareAcronym{iot}{
    short = IoT,
    long  = Internet of Things,
}
\DeclareAcronym{lr}{
    short = LR,
    long  = Linear Regression,
}
\DeclareAcronym{lstm}{
    short = LSTM,
    long  = Long Short-Term Memory,
}
\DeclareAcronym{maas}{
    short = MaaS,
    long  = Mobility as a Service,
}
\DeclareAcronym{mdln}{
    short = MDLN,
    long  = Multi-module Deep Learning Network,
}
\DeclareAcronym{ml}{
    short = ML,
    long  = Machine Learning,
}
\DeclareAcronym{mae}{
    short = MAE,
    long  = Mean Absolute Error,
}
\DeclareAcronym{nb}{
    short = NB,
    long  = Naive Bayesian,
}
\DeclareAcronym{olsr}{
    short = OLSR,
    long  = Ordinary Least-Squares Regression,
}
\DeclareAcronym{qr}{
    short = QR,
    long  = Quick Response,
}
\DeclareAcronym{rf}{
    short = RF,
    long  = Random Forest,
}
\DeclareAcronym{rmse}{
    short = RMSE,
    long  = Root Mean Square Error,
}
\DeclareAcronym{rnn}{
    short = RNN,
    long  = Recurrent Neural Networks,
}
\DeclareAcronym{sbc}{
    short = SBC,
    long  = Single-Board Computer,
}
\DeclareAcronym{smote}{
    short = SMOTE,
    long  = Synthetic Minority Over-sampling Technique,
}
\DeclareAcronym{sr}{
    short = SR,
    long  = Spatial Regression,
}
\DeclareAcronym{stgcn}{
    short = STGCN,
    long  = Spatio-Temporal Graph Convolutional Network,
}
\DeclareAcronym{uid}{
    short = UID,
    long  = Unique Identification Number,
}
\DeclareAcronym{upt}{
    short = UPT,
    long  = U-Park Token,
}
\DeclareAcronym{xgb}{
    short = XGB,
    long  = eXtreme Gradient Boosting,
}

\begin{document}

    \title{\LARGE \bf U-Park: A User-Centric Smart Parking Recommendation System for Electric Shared Micromobility Services}

    \author{Sen Yan, \textit{Graduate Student Member, IEEE}, Noel E. O'Connor, \textit{Member, IEEE}, \\and Mingming Liu, \textit{Senior Member, IEEE}
    	\thanks{S. Yan, N. E. O'Connor and M. Liu are with the SFI Insight Centre for Data Analytics and the School of Electronic Engineering, Dublin City University, Dublin, Ireland. This work is supported by Science Foundation Ireland under Grant No. \textit{21/FFP-P/10266} and \textit{SFI/12/RC/2289\_P2}. \textit{Corresponding author: Mingming Liu. Email: {\tt mingming.liu@dcu.ie}.}}}

    % make the title area
    \maketitle
    \thispagestyle{empty}
    \pagestyle{empty}

    \begin{abstract}\label{sec:abstract}
        Electric Shared Micromobility Services (ESMS) has become a vital element within the Mobility as a Service framework, contributing to sustainable transportation systems. However, existing ESMS face notable design challenges such as shortcomings in integration, transparency, and user-centred approaches, resulting in increased operational costs and decreased service quality. A key operational issue for ESMS revolves around parking, particularly ensuring the availability of parking spaces as users approach their destinations. For instance, a recent study illustrated that nearly 13\% of shared E-Bike users in Dublin, Ireland, encounter difficulties parking their E-Bikes due to inadequate planning and guidance. In response, we introduce U-Park, a user-centric smart parking recommendation system designed for ESMS, providing tailored recommendations to users by analysing their historical mobility data, trip trajectory, and parking space availability. We present the system architecture, implement it, and evaluate its performance using real-world data from an Irish-based shared E-Bike provider, MOBY Bikes. Our results illustrate U-Park's ability to predict a user's destination within a shared E-Bike system, achieving an approximate accuracy rate of over 97.60\%, all without requiring direct user input. Experiments have proven that this predictive capability empowers U-Park to suggest the optimal parking station to users based on the availability of predicted parking spaces, improving the probability of obtaining a parking spot by 24.91\% on average and 29.66\% on maximum when parking availability is limited.
    \end{abstract}

    % Note that keywords are not normally used for peer-review papers.

    \IEEEpeerreviewmaketitle

    \section*{Impact Statement}
        Shared electric micromobility services (ESMS), such as shared e-bikes, play a key role in sustainable urban mobility. However, ESMS often face parking challenges, affecting service quality and operating costs. U-Park, our user-centric smart parking recommendation system, solves this problem by accurately predicting the user's destination and suggesting the best parking space. With over 97.60\% prediction accuracy and a probability of obtaining a parking spot increased by 24.91\% on average when parking availability is limited, U-Park transforms ESMS, improving efficiency, reducing costs, and promoting sustainable urban mobility.
    
    \section{Introduction} 
        \label{sec: intro}
        With the rising interest in the shared economy and the increasing popularity of  \ac{em}, \ac{esms} have become a significant component of the \ac{maas} framework within modern intelligent transportation systems. \ac{em} options, such as \ac{ev}, \ac{ebike}, and \ac{escooter}, have garnered significant attention from the general public \cite{McQueen2020}. These modes of transportation demonstrate their potential advantages in terms of transportation convenience by addressing the last-mile transportation challenge in urban areas, environmental sustainability by replacing short trips typically made by private fossil fuel cars \cite{Cairns2017}, and promoting user health if used appropriately \cite{Gu2018}. Consequently, many European countries prioritise the adoption of \ac{esms} for sustainable, low-carbon transportation \cite{Astegiano2019, Galatoulas2020, Dias2021, Morfeldt2022}. For example, Ireland aims to expedite the adoption of \ac{ev}s to curb transport emissions\footnote{\url{https://www.gov.ie/en/publication/e62e0-electric-vehicle-policy-pathway}}.

Existing \ac{esms} face significant design challenges, notably the absence of service integration and user-centric approaches, which could impede their widespread adoption. Research reveals additional challenges facing \ac{esms}, such as improper parking behaviours \cite{Yan2022}, charging problems due to inadequate infrastructure \cite{Shui2020}, and accessibility challenges arising from fragmented planning and scheduling \cite{Chen2020}. To address these challenges, parking spot recommendation systems are being developed to aid users in locating suitable parking locations, thereby enhancing travel efficiency and mitigating parking costs. Specifically, sensor-based recommendation systems \cite{Tsai2021, Jabbar2021} usually rely on \ac{iot} infrastructure, offering improved information accuracy. However, these systems often fail to address the diverse and dynamic needs of users, resulting in limitations in flexibility and personalisation \cite{Nawara2020, Nawara2021}. Conversely, data-driven approaches, especially those based on \ac{ml} techniques, offer significant advantages in flexibility and adaptability to changing environments \cite{Teimoori2022, Chen2021}. Moreover, extensive studies \cite{Guidon2019, Zhou2022, Saum2020, Phithakkitnukooon2021, Xiao2020, Luo2021, Lee2022, Boonjubut2022, Liu2019, Jiang2019, Miao2021, Wang2021, Zong2019, Nawaz2020, Liang2022} have highlighted the potential of \ac{ml} algorithms and data-driven techniques to support recommendation systems for proactive and personalised mobility solutions, including destination prediction and parking availability prediction. However, to the best of the authors' knowledge, there is currently no smart parking solution integrating different components for \ac{esms} comprehensively.

This paper introduces U-Park, a user-centric smart parking recommendation system for \ac{esms} using \ac{ml} techniques. Our work builds upon prior research in this area \cite{Yan2022, Chen2021_1, Epperlein2018}. In \cite{Yan2022}, the parking behaviour of shared \ac{ebike} users was examined using a real-world dataset from Ireland. This study found that up to 12.9\% of users encountered difficulties parking correctly. In \cite{Chen2021_1}, an \ac{astgcn} was developed to predict bike availability at the start of user trips, achieving a \ac{mae} of 1.00. Lastly, in \cite{Epperlein2018}, a Bayesian classifier was devised for route prediction with Markov chains. This classifier can use a user's partial trajectory to update the posterior probabilities of the user's destination based on historical trip data. However, a key limitation of this work is that validation was performed solely with synthetic data, and it did not compare with advanced approaches like \ac{dl}-based methods.

To increase user satisfaction, encourage \ac{esms} adoption for the daily commute, and optimise system efficiency, U-Park introduces a novel approach based on \ac{ml} models to address the challenges of parking management within \ac{esms} by including the following key features:

\begin{itemize}
    \item {U-Park addresses users' parking needs in \ac{esms} proactively from the journey's start, eliminating the need to search for parking at the last minute.}
    \item {U-Park functions smoothly throughout all journey stages, minimising user input while remaining flexible enough to adapt to explicit instructions if provided.}
    \item {U-Park uses a user's current trajectory and historical mobility patterns to continually refine parking predictions, ensuring optimal accuracy and recommendations.}
\end{itemize}

The paper is structured as follows: In \autoref{sec: review}, we provide an overview of related work in existing \ac{esms} systems and \ac{ml} techniques applied. The research problem addressed in our study is detailed in \autoref{sec: model}. \autoref{sec: design} delves into the overall design of our enrollment system. The details of our prototype implementation are presented in \autoref{sec: implement}. \autoref{sec: result} presents our study results along with relevant discussions, and then the limitation of our current work is described in \autoref{sec: limitation}. Finally, we conclude our work in \autoref{sec: conclusion} and discuss future plans and improvements.
        
    \section{Literature Review} 
        \label{sec: review}
        In this paper, we consider the relevant literature from four perspectives: parking management solutions, current \ac{esms}, \ac{ml} methods in travel demand prediction, and \ac{ml} methods in trip destination prediction. 

\subsection{Parking Management Solutions}
    To address the challenges in parking management for shared mobility systems, many solutions have been investigated from various perspectives. These include optimising the layout of parking areas, examining the role of social norms, the impact of warning messages and monetary incentives, as well as developing systems for recommending parking spots.
    
    To optimise parking area layout, a collaborative optimisation model was proposed in \cite{Chen2024} using an improved genetic algorithm to minimise resident walking distances and enterprise costs. The designed model considered constraints including parking area coverage and bicycle allocation, ultimately optimising both the layout of shared bicycle parking areas and the number of bicycles deployed. On the other hand, the authors of \cite{Su2020} focused on the impact of social norms, warning messages, and monetary incentives, and conducted a comparative experiment using the control variable method. The results suggest that behavioural incentives are more effective than social norm intervention in shared bike parking management. 
    
    Some recommendation systems have also been designed for parking management by recommending parking locations based on sensor networks and \ac{iot} technologies \cite{Tsai2021, Jabbar2021}. As noted by the authors, this approach requires extensive infrastructure deployment, including sensors at intersections and parking lot entrances. However, these systems often struggle to meet the diverse and dynamic needs of users, as they tend to provide generic recommendations. Additionally, they typically require user interaction to initiate a parking search, which can be inconvenient for \ac{esms} users during their trips. In contrast, our system generates recommendations without relying on user input. By leveraging \ac{ml} algorithms and trip data collected from users, our recommendations take into account diverse user trip patterns and dynamic needs.
    
    In another previous study \cite{Liu2019}, a two-stage framework for destination prediction was proposed. This framework used hand-crafted rules, e.g., using several frequently-visited destinations in history as the candidate samples, to generate candidate destinations in the first stage and then employed a prediction model to predict the final destination from the candidate set, considering trip and customer attributes. 
    
    In contrast, our system, introduced in \autoref{sec: model}, relies entirely on a \ac{ml}-based model for trip destination prediction. Moreover, our system does not only rely on historical trip records but also incorporates trip trajectories. This means that our model not only provides a preliminary forecast based on historical data before the trip begins but also delivers in-journey predictions and recommendations. Furthermore, while the previous framework concluded with destination prediction, our approach integrates the destination as an input for the subsequent stage of the model.

\subsection{Current Electric Shared Micromobility Services} \label{subsec: system}
    \subsubsection{Improper Parking Behaviours}
        Recent research studies have extensively investigated inappropriate parking behaviours, with a focus on identification and mitigation \cite{Brown2020, Yan2022, Heinen2019, Gao2020, Chen2019}. For instance, in \cite{Brown2020}, data from five U.S. cities was analysed to detect instances of improper parking. Research in \cite{Heinen2019} demonstrated the correlation between improper parking behaviour and parking space availability, while \cite{B2022} proposed a real-time data-driven system employing \ac{ml} techniques to manage parking spaces, aiming to reduce shortages and conflicts among customers. However, these analyses and methods primarily operate at the trip conclusion and necessitate specific user-provided input data.

    \subsubsection{Pricing \& Incentive Policies}
        \ac{esms} employ diverse pricing structures, particularly pay-as-you-go models, with notable variations. In Dublin, Ireland, for example, MOBY\footnote{\url{https://www.dublincity.ie/residential/transportation/covid-mobility-measures/dublin-city-covid-19-mobility-programme/journey-planning}}, ESB\footnote{\url{https://www.irishtimes.com/business/2022/08/17/esb-launches-shared-ebike-scheme-across-dublin}}, and Tier\footnote{\url{https://irishcycle.com/2022/06/20/new-electric-bicycle-share-now-available-in-some-north-dublin-areas}} provide shared \ac{ebike}s, all featuring a consistent 1 euro starting fee but varying per-minute rates (5, 15, and 20 cents per minute, respectively). Similar pricing variations are observed in other regions as well\footnote{\url{https://www.expertreviews.co.uk/scooters/1416160/how-much-do-electric-scooters-cost-everything-you-need-to-know-whether-youre-buying}}. Moreover, \ac{esms} companies globally implement diverse incentive policies to promote vehicle repositioning. In a study \cite{Gao2020}, a reward of \textyen 1.43 (approximately \euro 0.20) per minute was proposed in China to encourage shared bike relocation. These policies consider multiple factors, such as travel distances and user preferences. Because of their practicality and proven effectiveness, we adopted similar mechanisms in our case study.

    \subsubsection{Relevant Datasets}
        In recent years, numerous research studies have delved into the realm of \ac{esms}. However, due to the limited accessibility of data directly from \ac{esms} providers, research and exploration in this domain have been restricted. Some studies have introduced datasets with similar characteristics\cite{Guidon2019, Zhou2022, Saum2020, Phithakkitnukooon2021}. These datasets include trip history data, such as aperture and arrival times and locations, trip duration, distance, and trip trajectory data with timestamps, \ac{gps} coordinates, and vehicle indices, as seen in \cite{Zhou2022}. Nonetheless, a common limitation across these datasets is the absence of user-related features, which hinders the provision of personalised services. Therefore, after careful consideration, we opted to utilise the MOBY dataset. Although it has a limited volume, it contains information about anonymous users, enabling us to address the aforementioned challenges effectively.
    
    In summary, addressing improper parking in \ac{esms} systems \cite{Yan2022} is important, and this is highly influenced by parking space availability \cite{Heinen2019}. In the next sections, we examine \ac{ml} techniques used in \ac{esms} for trip demand and destination prediction. The latter helps predict the endpoint of a current journey, prompting users to park correctly, while the former represents the number of departures and arrivals in an area or station, determining parking space availability.

\subsection{\ac{ml} Methods in Travel Demand Prediction} \label{subsec: ml-demand}
    Hourly demand forecasting for \ac{esms} has been addressed using various \ac{ml} algorithms and models in different cities globally. In this section, we provide an overview of previous methods and techniques for hourly demand prediction.
    
    In previous studies \cite{Guidon2019, Zhou2022, Saum2020}, classic \ac{ml} methods like \ac{sr} and \ac{lr} have been employed to forecast hourly demand for shared \ac{ebike}s. In \cite{Zhou2022}, for example, the authors focused on predicting \ac{ebike} travel demand. The proposed method, \ac{gtwr}, achieved a significantly lower relative error of -1.07\% compared to 7.85\% for \ac{olsr} and -1.43\% for \ac{gwr}, demonstrating its superior predictive performance.
    
    Recently, \ac{dl} techniques, including \ac{lstm} in \ac{rnn}, \ac{cnn}, and \ac{gcn}, have gained significant attention in micromobility research \cite{Lee2022, Phithakkitnukooon2021, Luo2021, Xiao2020, Chen2021}. These models are adept at capturing temporal patterns. For instance, in \cite{Lee2022}, \ac{convlstm} was employed to predict next-hour bicycle demand using rental histories alongside meteorological data. The use of \ac{gcn}-based methods is also on the rise in this prediction task, given their ability to model non-Euclidean data localities \cite{Luo2021}. In \cite{Xiao2020}, bike pickup and return demand were predicted using the \ac{stgcn} model, which outperformed three baseline models, \ac{rnn}, \ac{lstm}, and \ac{gru}, in terms of prediction accuracy and computational efficiency. Inspired by cognitive attention in \ac{dl}, attention mechanisms have been employed to enhance the prediction performance of \ac{gcn} models. In \cite{Chen2021}, \ac{astgcn} was proposed and applied to predict the bike availability at each bike parking station. This approach exhibited superior performance when compared to other existing methods, including \ac{stgcn} and \ac{xgb}, when tested on real-world datasets. Our present research builds upon these concepts by adopting the method proposed in \cite{Chen2021} to design the parking space availability prediction module within the \ac{esms} architecture.
        
    Finally, extensive research has highlighted that prediction accuracy can be affected by various journey attributes, including spatial attributes (e.g., origin and destination), temporal attributes (e.g., hour of the day and day of the week), and weather conditions (e.g., temperature and wind speed) \cite{Guidon2019, Zhou2022, Lee2022, Xiao2020, Luo2021}. Consequently, these journey attributes have also been included in the \ac{ml}-based method used in this study. 
    
\subsection{\ac{ml} Methods in Trip Destination Prediction} \label{subsec: ml-destination}
    Methods and models used in trip destination prediction in relevant papers are 
    % summarised in \autoref{table: destination review}. Selected works are 
    detailed below. Generally, travel destination prediction can be categorised based on the data utilised into two types: history-based and \ac{gps}-based. The former one focuses on the prediction based on the trip records, including the temporal and spatial features of only the origins and destinations, while the latter one aims to predict the destination based on time-series data reflecting the trip trajectory.

    \subsubsection{Trip History-Based Prediction}
    
        Trip history datasets provide temporal and spatial information about the origin and destination of each journey. Researchers have employed various techniques, including \ac{arima}, \ac{convlstm}, \ac{cnn}, \ac{gcn}, to predict destination using trip history data \cite{Miao2021, Jiang2019, Liu2019}. For instance, a \ac{dpnst} was introduced in \cite{Jiang2019}, integrating \ac{lstm}, \ac{cnn} and a Fully Connected Neural Network, each extracting user behaviour, spatial features, and external features like weather conditions, respectively. The results were compared to alternative models, including \ac{lstm} and \ac{nb} \cite{Epperlein2018}.
        
        % As demonstrated in \autoref{table: destination review}, 
        Our review illustrated that complex models with strong performance are adaptable to large datasets. However, considering the limitations of our trip history dataset, we chose \ac{ann} in our study because of its flexibility, adaptability, and suitability for handling small datasets, making it an ideal choice for our research.
    
    \subsubsection{Trajectory-Based Prediction}
        In parallel, numerous research studies have explored destination prediction based on \ac{gps} data for various applications \cite{Nawaz2020, Zong2019, Liang2022}. For instance,  in \cite{Liang2022}, the authors proposed an \ac{mdln} to solve the problem of destination prediction. As the journey trajectory is a temporally-ordered \ac{gps} location collection, \ac{rnn} \cite{Rossi2020} and \ac{lstm} \cite{Nawaz2020} are also widely used in trajectory-based destination prediction thanks to their ability to handle temporal and sequential data effectively. Besides, recent research has indicated that the initial $k$ and final $k$ segments of a \ac{gps} trajectory significantly contribute to destination prediction \cite{Lv2020, Liang2022, Tang2021}, so we adopted the same method. 
        
To sum up, \ac{ml} models for both hourly demand and destination prediction commonly incorporate external factors such as weather and points of interest in addition to spatial and temporal information related to trip origins and destinations \cite{Zhou2022, Boonjubut2022, Jiang2019, Miao2021}. \ac{rnn} models play a crucial role in these tasks due to their effectiveness in time-series prediction problems \cite{Phithakkitnukooon2021, Xiao2020, Luo2021, Lee2022, Boonjubut2022, Jiang2019, Liu2019, Miao2021, Nawaz2020, Liang2022}. Considering our dataset size, the model complexity, and prediction accuracy, we decided to employ a simple \ac{ann} for initial destination prediction based on trip history at the start of a journey, while using an \ac{rnn} model during the trip based on \ac{gps} data in our study.

    \section{System Model}
        \label{sec: model}
        In this section, we introduce our research problem and provide an in-depth explanation of each question. To enhance clarity, we use a shared \ac{ebike} system as an example of \ac{esms} systems, but U-Park applies to other \ac{esms} systems as well. We segment the user journey into three stages: pre-journey (from the moment users decide to start a trip and scan the \ac{qr} code attached to the \ac{ebike} until they are unlocked), in-journey (from the point of unlocking the bike until users are near their destination and ready to end the trip), and post-journey (from the moment users are near the end of the trip until they receive final recommendations from U-Park, lock up the bike, and complete the payment process). Specifically, we consider the trip to be concluding if the user's current position is within a certain distance, e.g., $dis$ metres, of the predicted destination.

\subsection{Overall Research Problem}
    \textit{Notations:} In essence, U-Park's primary objective is to solve improper parking behaviours in \ac{esms} by providing parking station recommendations as a trip nears completion. Given this context, we can formulate the parking station recommendation problem. With an \ac{esms} system, the operator has established a total of $S$ predefined parking stations. We introduce the set of parking stations, denoted as $\mathbb{S}$, as follows:

    \begin{equation} \label{equ: station definition}
        \mathbb{S} = \{s_1, s_2, \dots, s_S\}
    \end{equation}
    
    For a given user, their historical record $\mathbb{H}$ and \ac{gps} record $\mathbb{G}$ contain information about a total of $N$ past trips within the \ac{esms}. Leveraging $\mathbb{H}$ and $\mathbb{G}$, our goal is to devise a system to make recommendations for suitable parking stations near the destination of the user's ongoing trip. To achieve this, we use $\mathbb{T}_i$ to represent their $i^{th}$ trip. For each trip $\mathbb{T}_i$, we define its \ac{gps} record $\mathbb{T}_i^G$ and historical record $\mathbb{T}_i^H$ as follows.
    
    Firstly, let us define a \ac{gps} record of a single trip $\mathbb{T}_i^G$ by

    \begin{equation} \label{equ: trip GPS definition}
        \mathbb{T}_i^G = [(t_1^i, p_1^i); (t_2^i, p_2^i); \dots; (t_{L_i}^i, p_{L_i}^i)]
    \end{equation}

    \noindent where the sequences $(t_1^i, \dots, t_{L_i}^i)$ and $(p_1^i, \dots, p_{L_i}^i)$ represent lists of timestamps and \ac{gps} positions, both having the same length denoted as $L_i$. Thus, the size of $\mathbb{T}_i^G$ is $L_i \times 2$. The initial position $p_1^i$ and the final position $p_{L_i}^i$ correspond to the origin and destination of the trip $\mathbb{T}_i^G$, while the first timestamp $t_1^i$ and last timestamp $t_{L_i}^i$ represent its departure time and arrival time, respectively. Given the context above, the user's historical record for a single trip $\mathbb{T}_i^H$ is defined as:

    \begin{equation} \label{equ: trip history definition}
        \mathbb{T}_i^H = [(t_1^i, p_1^i, t_{L_i}^i, p_{L_i}^i)]
    \end{equation}

    \noindent which means the history of one specific trip $\mathbb{T}_i^H$ contains the spatial and temporal characteristics of its origin and destination and the size of $\mathbb{T}_i^H$ is $1 \times 4$. Expanding on the context given above, the user's historical record for $N$ trips is:

    \begin{equation} \label{equ:  history record definition}
        \mathbb{H} = \{\mathbb{T}_1^H, \mathbb{T}_2^H, \dots, \mathbb{T}_N^H\}
    \end{equation}
    
    \noindent and similarly, the user's \ac{gps} record for $N$ trips is:

    \begin{equation} \label{equ: GPS report definition}
        \mathbb{G} = \{\mathbb{T}_1^G, \mathbb{T}_2^G, \dots, \mathbb{T}_N^G\}
    \end{equation}

    \noindent Finally, based on the list of all trip destinations within this user's \ac{gps} records, denoted as $(p_{L_1}^1, p_{L_2}^2, \dots, p_{L_N}^N)$, we define the set of destinations for this user as follows: 

    \begin{equation} \label{equ: destination set}
        \mathbb{D} = \{d_1, d_2, \dots, d_D\}
    \end{equation}

    \noindent where any $d_j \in \mathbb{D}$ represents a unique destination, and $D$ is the length of $\mathbb{D}$, so for any given trip $\mathbb{T}_i$ in $\mathbb{G}$, we have the destination $p_{L_i}^i \in \mathbb{D}$.

    \textit{Problem:} For an ongoing trip $\mathbb{T}_k = [(t_1^k, p_1^k); \dots)]$, the primary goal of our U-Park system is to determine a suitable parking station $\hat{s_k} \in \mathbb{S}$ and make recommendations aiming at increasing the user's chance to obtain an available parking space near the predicted destination $\hat{d_k} \in \mathbb{D}$, which is predicted by \ac{ml} models within U-Park system.

    \textit{Remark:} The $\mathbb{S}$ and $\mathbb{D}$ defined in this section represent different sets. The former is the predefined station set, while the latter is the set of destinations from this user's previous trips. The destinations may or may not be included in $\mathbb{S}$. Only for new users without history records, we specify that $\mathbb{D} = \mathbb{S}$. To address our primary research question, we divide it into three \ac{ml} tasks corresponding to the three stages introduced earlier. We define these tasks as follows and provide detailed explanations in the subsequent sections: (i) history-based destination prediction in the pre-journey stage; (ii) trajectory-based destination prediction in the in-journey stage; and (iii) parking space availability prediction in the post-journey stage.

\subsection{History-Based Destination Prediction}
    \textit{Notations:} As introduced above, this history-based prediction model is applied at the pre-journey stage to predict the destination of this trip. In addition to the departure and arrival timestamp and position introduced in \eqref{equ: trip history definition}, each specific trip $\mathbb{T}_k$ includes a set of extra features $f_k^H$, such as temporal features (e.g., day of the week) and weather conditions (e.g., temperature). Denoting $N_f^H$ as the number of $f_k^H$, the length of the extended historical record for $\mathbb{T}_k$ becomes $M_H = N_f^H + 4$. In other words, the size of the extended historical record for $N$ trips in total amounts $N \times M_H$.

    \textit{Problem:} When dealing with an upcoming trip $\mathbb{T}_k = [(t_1^k, p_1^k, f_k^H)]$, the research question at this stage is to solve the classification task of determining the category within $\mathbb{D}$ that the trip $\mathbb{T}_k$ belongs to, or in other words, to identify a $d_H^k \in \mathbb{D}$ as the destination for $\mathbb{T}_k$.

    \textit{Remark:} This problem is solved every time a user gains authorisation to unlock a vehicle in this \ac{esms}. Since the model only relies on the user's historical record and external features, the prediction result for the current trip $\mathbb{T}_k$ remains constant until the dataset is enriched by new records.

\subsection{Trajectory-Based Destination Prediction}
    \textit{Notations:} As previously mentioned, this trajectory-based prediction model is employed during the in-journey stage to improve the prediction result made at the pre-journey stage. In addition to the timestamp and \ac{gps} position sequences in \eqref{equ: trip GPS definition}, each specific trip $\mathbb{T}_k$ with a length of $L_k$ incorporates supplementary features $f_k^G$, such as temporal features (e.g., day of the week) and weather conditions (e.g., temperature) obtained for each timestamp. Denoting $N_f^G$ as the number of $f_k^G$, the dimensions of the extended \ac{gps} record for $\mathbb{T}_k$ can be represented as $L_k \times M_G$, where $M_G = N_f^G + 2$ denotes the number of features for each timestamp. 

    \textit{Problem:} At this stage, the research question for this ongoing trip $\mathbb{T}_k = [(t_1^k, p_1^k, f_k^G); (t_2^k, p_2^k, f_k^G); \dots]$ is to solve the classification task of determining the category within $\mathbb{D}$ that the trip $\mathbb{T}_k$ corresponds to, or in other words, to designate $d_G^k \in \mathbb{D}$ as the destination for $\mathbb{T}_k$.

    \textit{Remark:} Concerning \ac{gdpr}, \ac{gps} data will be converted and then integrated into our system rather than being directly input. The processing of \ac{gps} data will be introduced in \autoref{sec: implement}. This problem is resolved each time a new \ac{gps} position is observed during the current trip. Since this model relies on external features and the user's real-time trajectory, it is feasible to update the prediction result when a new \ac{gps} point is included. 

\subsection{Parking Space Availability Prediction}
    \textit{Notations:} Lastly, we articulate the parking space availability prediction problem as follows. Let $A_t^k$ denote the number of available parking docks or spaces at station $s_k \in \mathbb{S}$ at time $t$. $\textbf{A}_t$ represents the vector consisting of the numbers of the available parking docks or spaces at all $S$ stations in total at time $t$. Consequently, the length of $\textbf{A}_t$ is $S$. Moreover, we use $f_t^k$ to denote the value of external features at station $s_k$ at time $t$, such as temporal features (e.g., day of the week) and weather conditions (e.g., temperature) for model training, and we use $M_F$ to represent the length of $f_t^k$. Similarly, we denote the feature set values of all parking stations at time $t$ by  $\textbf{f}_t$, so the size of $\textbf{f}_t$ is $S \times M_F$.
    
    \textit{Problem:} Mathematically, the learning objective of this prediction model, which is applied in the post-journey stage, is to find a function $\textbf{H}(.)$ that addresses the following problem:
    
    \begin{equation}
        \textbf{A}_{t+1:t+n} = \textbf{H}(\textbf{A}_{t-m+1:t};\textbf{f}_{t-m+1:t})
    \end{equation}
    
    \noindent where $m$ and $n$ represent the input and output lengths of this model, respectively. Besides, $t+1:t+n$ denotes the output as a sequence of vectors from time $t+1$ to $t+n$.

    \textit{Remark:} Predictions for this problem persist until the user confirms the destination of this trip or reaches locations with the highest prediction occurrences. This model is influenced by the arrival time and external features, and the arrival time can be obtained using any commonly used map \ac{api}.
    
    To explain the reason for making multiple prediction results, we consider the following example: A user plans to take a shared \ac{ebike} trip from \textit{Point A} to a crowded bike station \textit{Point C}, which is close to a less crowded station \textit{Point B}. Using multiple prediction results for each bike station in the next hour, at certain intervals (e.g., 15 minutes), U-Park can forecast that \textit{Point C} might have fewer parking spaces available than \textit{Point B} upon arrival with high confidence. In response, U-park will recommend the user to park at \textit{Point B} and walk to \textit{Point C} rather than park at \textit{Point C} to avoid potential time costs in looking for parking spaces. This provides a more convenient solution and helps prevent wasted time searching for parking spots. As a result, users can better deal with uncertainty arising during their travel, improving overall experience and satisfaction.
    
    \section{System Architecture} 
        \label{sec: design}
\begin{figure*}[htbp]
    \vspace{-0.1in}
    \centering
    \includegraphics[width=\linewidth]{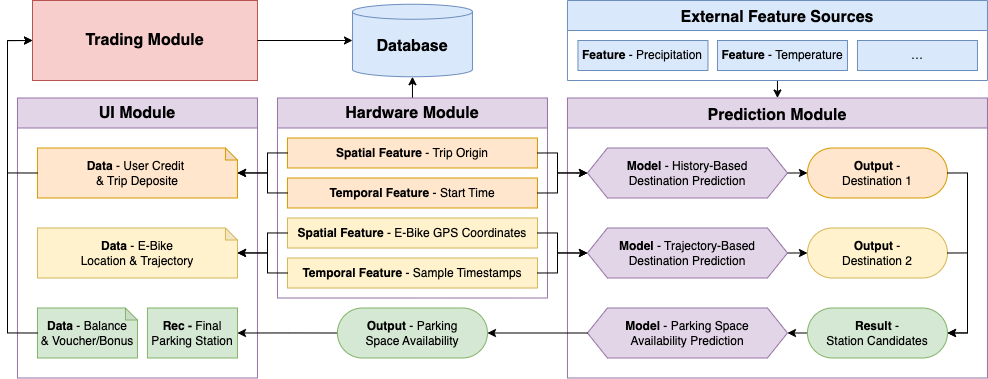}
    \caption{The proposed system architecture for U-Park.}
    \label{fig: overview}
    \vspace{-0.2in}
\end{figure*}

        This section presents the proposed framework structure, U-Park, to address the research questions discussed in \autoref{sec: model}. An overview of the system architecture is provided in \autoref{fig: overview}, with hardware and user interface modules involved in all three journey stages: pre-journey, in-journey, and post-journey, as defined in the previous section. Again, we use a shared \ac{ebike} system as an example of \ac{esms} systems for clarity.

U-Park consists of three main modules: a hardware module (e.g., \ac{ebike}s with various sensors), a user interface module (e.g., a mobile application), and a prediction module based on \ac{ml} prediction algorithms. Additionally, it incorporates external feature sources, a database, and a trading module to manage financial transactions.

\subsection{Hardware Module} \label{subsec: arc-pi}
    The hardware module attached to U-Park's \ac{ebike}s serves three primary functions: collecting physical data from the bikes, calculating trip deposits, and controlling motor rotation. This module can be built on a \ac{sbc}, e.g., a Raspberry Pi with various sensors connected to interact with the cloud database. When an unlocking or locking request is received, the \ac{sbc} will unlock or lock the \ac{ebike} accordingly and calculate the trip deposit, and the motor status will be stored in the database and updated in real-time throughout this trip. Additionally, this module will upload the \ac{gps} coordinates of \ac{ebike}s to our database, allowing both administrators and users to track the current locations of \ac{ebike}s.
    
\subsection{User Interface Module} \label{subsec: arc-app}
    An Android mobile application has been developed to interact with users. It collects trip origin data, communicates with the prediction model, and delivers recommendations back to users. The application handles user credits, payments, and data collection, which is then uploaded to the database through the trading module. Furthermore, the application employs map \ac{api}s and \ac{qr} code scanning to enable users to view their current position, locate nearby \ac{ebike}s, and unlock available \ac{ebike}s by scanning \ac{qr} codes.
    
    When users open and log into the mobile application, their positions are located on the map according to \ac{gps} coordinates. The map also displays the locations and availability of \ac{ebike}s using colour-coded markers, e.g., in \autoref{fig: mobile-app-parking-zones}. Users can input their intended destination $d_U^k$ for a trip $\mathbb{T}_k$. Once the user provides a destination, the system proceeds directly to the post-journey stage, as described in \autoref{sec: model}. However, our primary focus is on scenarios where users may not provide input to the system, allowing us to address more complex and versatile situations.
    
    Users can scan the \ac{ebike}'s \ac{qr} code to start the credit validation and deposit payment process when clicking on an available \ac{ebike} marker. The deposit amount is determined by the total duration from the origin to the predicted destination, as determined by the history-based model in \autoref{sec: model}. The equation employed in this step \eqref{equ: deposit} is detailed in the following section. If the user has sufficient remaining credit to cover the deposit, the device is unlocked after payment, and the journey status is updated in our database. Otherwise, the user is advised to add credit to their account and try again.
    
    The \ac{ebike} \ac{gps} is continuously captured and uploaded until the user ends the trip. Upon trip completion, the balance amount $F_{balance}$ is calculated by \eqref{equ: balance}, which is determined by the trip duration, the minimum distance to predefined stations, and the deposit paid at the start of this trip.
    
    \begin{equation} \label{equ: balance}
        F_{balance} = f \cdot (t_{L_k}^k - t_1^k) + F_{parking} - F_{deposit}
    \end{equation}
    
    In \eqref{equ: balance}, $f$ represents the charge per second, $t_{L_k}^k$ is the actual arrival timestamp at the destination $d_k$ and $F_{parking}$ is the predefined parking fee defined in \autoref{sec: implement}. If $F_{balance} \geq 0$, it indicates that the user needs to be charged again at the end of the trip, and if $F_{balance} < 0$, a refund is provided.

\subsection{Prediction Module} \label{subsec: arc-model}
    The prediction module serves two purposes: destination prediction and parking space availability prediction. We employ a two-step prediction model to improve the accuracy of destination prediction. When users initiate a trip, we first predict a list of potential destinations, i.e., a set of $d_H^k$, using our history-based prediction model. Subsequently, we refine it by our trajectory-based prediction module result, i.e., $d_G^k$. In other words, destination prediction is divided into two stages: pre-journey and in-journey. The final estimated trip destination $\hat{d_k}$ is determined by both $d_H^k$ and $d_G^k$.
    
\begin{figure*}[htbp]
        \vspace{-0.1in}
        \centering
        \includegraphics[width=\linewidth]{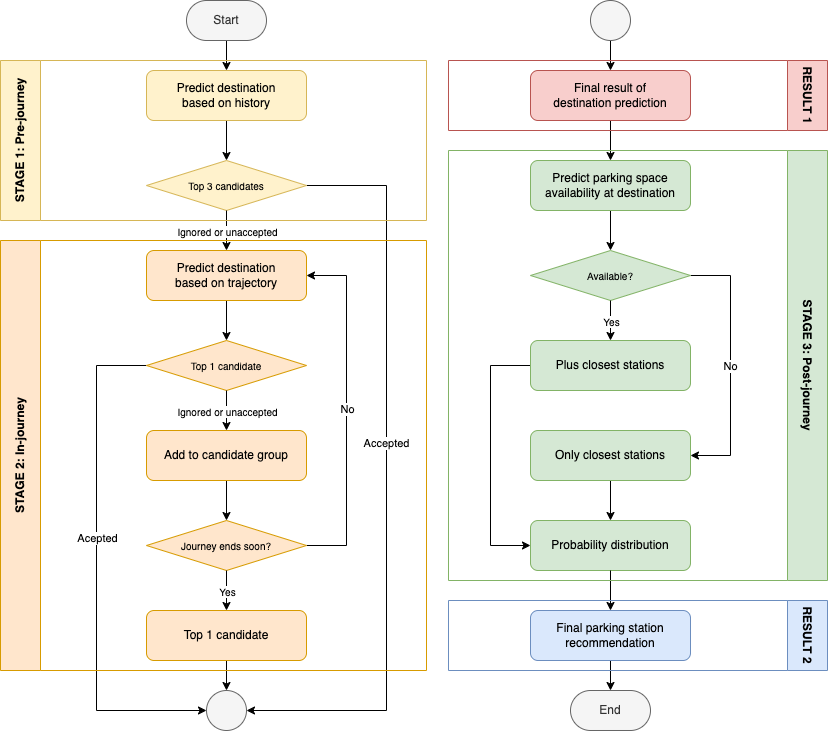}
        \caption{Prediction module workflow.}
        \label{fig: prediction module flowchart}
        \vspace{-0.2in}
    \end{figure*}

    The workflow of the entire prediction module is illustrated in \autoref{fig: prediction module flowchart}, where items coloured in yellow, orange, and green represent history-based and trajectory-based destination predictions as well as the parking space availability prediction, respectively. The red item summarises the final result of the destination prediction task, while the blue items calculate probability distributions to make a recommendation regarding which parking station to use. For clarity, we summarise the notations used for destination prediction in \autoref{table: destination notations}.
    
    \begin{table}[htbp]
        \caption{Notations for destination prediction tasks.}
        \vspace{-0.1in}
        \label{table: destination notations}
        \begin{tabularx}{\linewidth}{@{\extracolsep{\fill}}c X}
            \toprule
            Notation    &   Explanation \\  
            \midrule
            $d_k$       &   the actual destination of trip $\mathbb{T}_k$ which will be recorded when the journey ends \\
            $d_H^k$     &   the predicted destination of trip $\mathbb{T}_k$ which results from our history-based model \\
            $d_G^k$     &   the predicted destination of trip $\mathbb{T}_k$ which results from our trajectory-based model \\
            $d_U^k$     &   the user-planned destination of trip $\mathbb{T}_k$ which is determined by the user when the journey starts \\
            $\hat{d_k}$ &   the overall result of the destination prediction module which is determined by $d_H^k$ and $d_G^k$ \\
            $\hat{s_k}$ &   the overall result of U-Park, the final parking station recommended to the user when $\hat{d_k}$ is decided \\
            \bottomrule
        \end{tabularx}
        \vspace{-0.1in}
    \end{table}
    
    \subsubsection{Pre-journey} \label{subsubsec: arc-model-pre}
        Our prediction module focuses on destination prediction based on the user's history at this stage. Our classification model, as introduced in \autoref{sec: model}, is a probabilistic classifier, which results in a probability distribution over a set of categories instead of the most likely category. Consequently, we consider multiple destination candidates with the highest probabilities and calculate the trip deposit $F_{deposit}$ accordingly.
        
        Following the notations, when a user starts a trip $\mathbb{T}_k$ from the origin $p_1^k$, U-Park collects necessary data, including the departure location, starting time, weather conditions, and battery remaining power. Temporal, spatial, and weather-related data are used for destination prediction, while the remaining battery power is compared with the expected travel distance to determine if it is sufficient to complete this trip. If not, U-Park helps the user find other suitable \ac{ebike}s. The predicted destination candidates $\mathbb{D}_H^k$, represented as a set of $d_H^k$, are then recommended for user confirmation. Depending on the user responses, two cases are considered in this step.
        
\begin{figure}[htbp]
            \vspace{-0.1in}
            \centering
            \subfigure[Pre-journey recommendation.]{
                \includegraphics[width=0.22\textwidth]{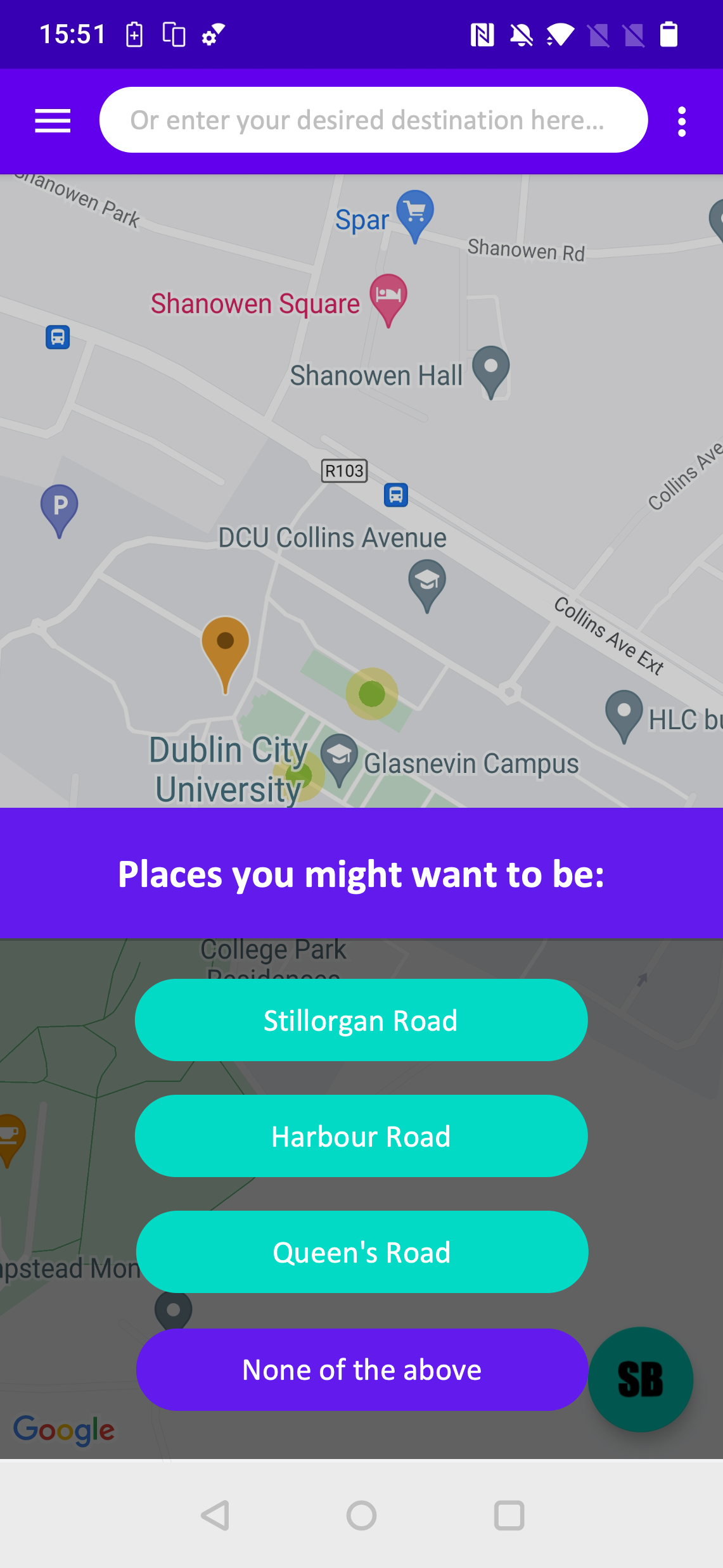}
                \label{subfig: pre-rec}
            }
            \subfigure[In-journey recommendation.]{
                \includegraphics[width=0.22\textwidth]{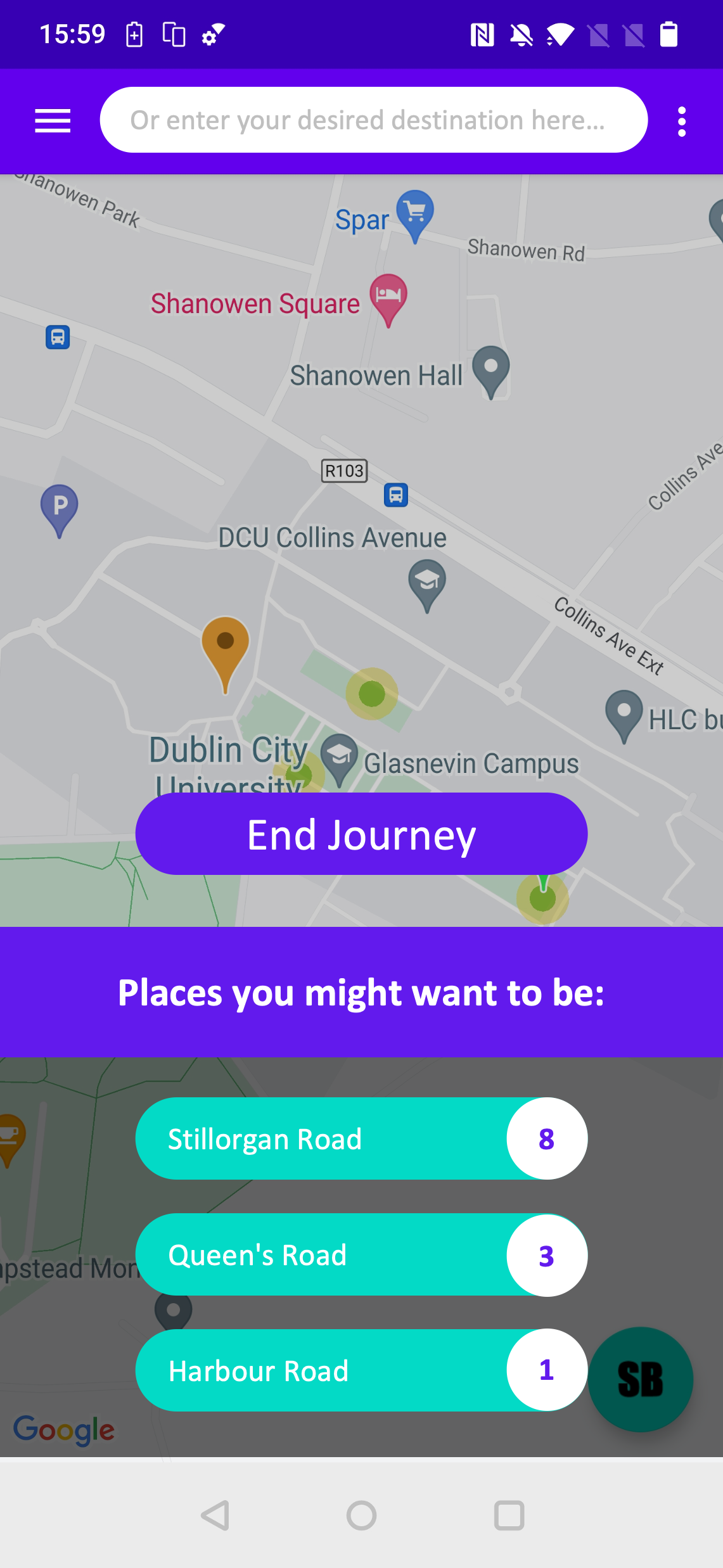}
                \label{subfig: in-rec}
            }
            \vspace{-0.1in}
            \caption{Sample user interfaces of recommendations.}
            \label{fig: sample UI}
            \vspace{-0.1in}
        \end{figure}

        \textit{Case 1:} If any result $d_x$ in $\mathbb{D}_H^k$ matches $d_U^k$, i.e., $\exists d_x \in \mathbb{D}_H^k$ $d_x = d_U^k$, and the users confirm it by selecting it from the list given in our mobile application, or the users enter a destination in the input box as shown in \autoref{subfig: pre-rec}, we denote this case as $d_U^k \in \mathbb{D}_H^k$. In this case, our prediction module adopts $d_U^k$ as the final destination result, i.e., $\hat{d}_k \gets d_U^k$, and calculates the deposit accordingly. Then, the system proceeds directly to the post-journey stage. Mathematically, the deposit for trip $\mathbb{T}_k$ is determined by the timestamp $t_U^k$ when the user arrives at $d_H^k$, estimated by maps \ac{api}.

        \textit{Case 2:} If none of the predicted results in $\mathbb{D}_H^k$ matches $d_U^k$, i.e., $\forall d_x \in \mathbb{D}_H^k$ $d_x \neq d_U^k$, or if the user disregards the recommendation in this stage, we denote this case as $d_U^k \notin \mathbb{D}_H^k$. In this scenario, the system estimates the expected deposit value based on the user's historical destinations $\mathbb{D}$ and proceeds to the in-journey stage. To estimate the expected deposit, we define $\mathbb{D}_k$ as the set of destinations for all trips in the user's historical record starting from the origin $p_1^k$, so $\mathbb{D}_k \subset \mathbb{D}$. Let $D_k$ denote the number of elements in $\mathbb{D}_{p_1^k}$, so we have $D_k \leq D$. For any destination $d_l^k \in \mathbb{D}_k$, the arrival timestamp is represented by $t_l^k$. We use $N_l$ to denote the total number of trips starting at $p_1^k$, and $N_l^k$ to denote the number of trips starting at $p_1^k$ and ending at $d_l^k$. Therefore, the weight $w_l^k$ of destination $d_l^k$ is calculated as:

        \begin{equation} \label{equ: probability}
            w_l^k = \frac {N_l^k}{N_l} \qquad
        \end{equation}

        \noindent where $\sum_{l=1}^{D_k}w_l^k = 1$, so the deposit for trip $\mathbb{T}_k$ is determined by the timestamps of arrival at all destinations in $\mathbb{D}_k$ and their corresponding weights.

        To sum up, the equation for deposit calculation is given by

        \begin{equation} \label{equ: deposit}
            F_{deposit} = 
            \begin{cases}
                f \cdot (t_U^k - t_1^k)  & \text{if $d_U^k \in {\mathbb{D}^k_H}$} \\
                f \cdot \sum_{l=1}^{D_k} w_l^k \cdot (t_l^k - t_1^k)  & \text{if $d_U^k \notin {\mathbb{D}^k_H}$}
            \end{cases}
        \end{equation}
        
        \noindent where $f$ is the same constant value used in \eqref{equ: balance} representing the charge per second.
    
    \subsubsection{In-journey} \label{subsubsec: arc-model-in}
        
        During this stage, the continuous \ac{gps} coordinates of the \ac{ebike} are used to predict the trip destination. Predicted trip destinations $d_G^k$ are recommended to the user at specific intervals and saved in a sorted list $\mathbb{D}_G^k$, based on the frequency of occurrences. Users can customise recommendation frequency, confirm destinations at any time, or choose to ignore the recommendations while cycling. When users decide to end their journey and make a payment, they click the corresponding button in the mobile app. Three cases need to be considered in this step, depending on user responses.

        \textit{Case 1:} If any predicted result $d_y$ in $\mathbb{D}_G^k$ matches $d_U^k$, i.e., $\exists d_y \in \mathbb{D}_G^k$ $d_y = d_U^k$, and the user selects it from the list in our mobile application, as shown in \autoref{subfig: in-rec}, we denote this case as $d_U^k \in \mathbb{D}_G^k$. In this case, our prediction module adopts $d_U^k$ as the final result, i.e., $\hat{d}_k \gets d_U^k$. The system enters the post-journey stage without further destination prediction.

        \textit{Case 2:} If the user ignores all the recommendations and is not ready to end the trip, we denote this case as $\nexists d_U^k$. The system will use the result at the top of $\mathbb{D}_G^k$, which is the prediction with the highest occurrence value, e.g., a set of predicted destinations followed by the number of their occurrence in \autoref{subfig: in-rec}. Then, U-Park will begin the post-journey stage but continue tracking user responses to the predictions until an end command is issued.

        \textit{Case 3:} If the user does not choose any predicted destination but is ready to end the trip by clicking the ``End Journey" button shown in \autoref{subfig: in-rec}, we denote this case as $d_U^k \notin \mathbb{D}_G^k$, and set the user's current location $p_{-1}^k$ as the final result, i.e., $\hat{d}_k \gets p_{-1}^k$. The system will enter the post-journey stage.

        \autoref{alg: des determine} demonstrates the workflow of our destination prediction module, in which HistoryBasedModel(.) and TrajectoryBasedModel(.) represent the prediction models introduced earlier in this section. They are used to compute the predicted destination results based on trip history and trip trajectory, respectively. Specifically, if a user enters a desired destination, as shown in \autoref{alg: des determine}, U-Park will adopt user input as the final destination result, skip destination prediction, and proceed directly to the post-journey stage.

        \begin{algorithm}[ht]
            \caption{Workflow of destination predictions}
            \label{alg: des determine}
            \begin{algorithmic}
                \renewcommand{\algorithmicrequire}{\textbf{Input:}}
                \renewcommand{\algorithmicensure}{\textbf{Output:}}
                \REQUIRE An ongoing trip $\mathbb{T}_k = [(t_1^k, p_1^k); \dots]$, extra features $f_k$, distance threshold $dis$, (user-planned destination $d_U^k$) 
                \ENSURE  Prediction of destination $\hat{d_k}$
                \IF{$d_U^k$ is initially given}
                    \STATE $\hat{d_k} \gets d_U^k$
                    \RETURN $\hat{d_k}$
                \ENDIF
                \WHILE{length$(\mathbb{T}_k) = 1$}
                    \STATE $\mathbb{D}_H^k \gets $ \hyperref[subsubsec: arc-model-pre]{HistoryBasedModel}$(\mathbb{T}_k)$
                    \IF{$\exists d_x \in \mathbb{D}_H^k$ $d_x = d_U^k$ (i.e., $d_U^k \in \mathbb{D}_H^k$)}
                        \STATE $\hat{d_k} \gets d_x$
                        \RETURN $\hat{d_k}$
                    \ENDIF
                    \STATE wait for more trip data to be added 
                \ENDWHILE
                \WHILE{(UserStop$(\mathbb{T}_k)=False$) \& ($\hat{d_k}=nothing$)}
                    \STATE $\mathbb{D}_G^k \gets$ \hyperref[subsubsec: arc-model-in]{TrajectoryBasedModel}($\mathbb{T}_k$)
                    \IF{$\exists d_y \in \mathbb{D}_G^k$ $d_y = d_U^k$}
                        \STATE $\hat{d_k} \gets d_y$
                        \RETURN $\hat{d_k}$
                    \ELSIF{UserStop$(\mathbb{T}_k)=True$}
                        \STATE $\hat{d_k} \gets p_{-1}^k$
                        \RETURN $\hat{d_k}$
                    \ENDIF
                \ENDWHILE
            \end{algorithmic}
        \end{algorithm}
        
    \subsubsection{Post-journey} \label{subsubsec: arc-model-post}
\begin{figure}[htbp]
            \vspace{-0.1in}
            \centering
            \includegraphics[width=\linewidth]{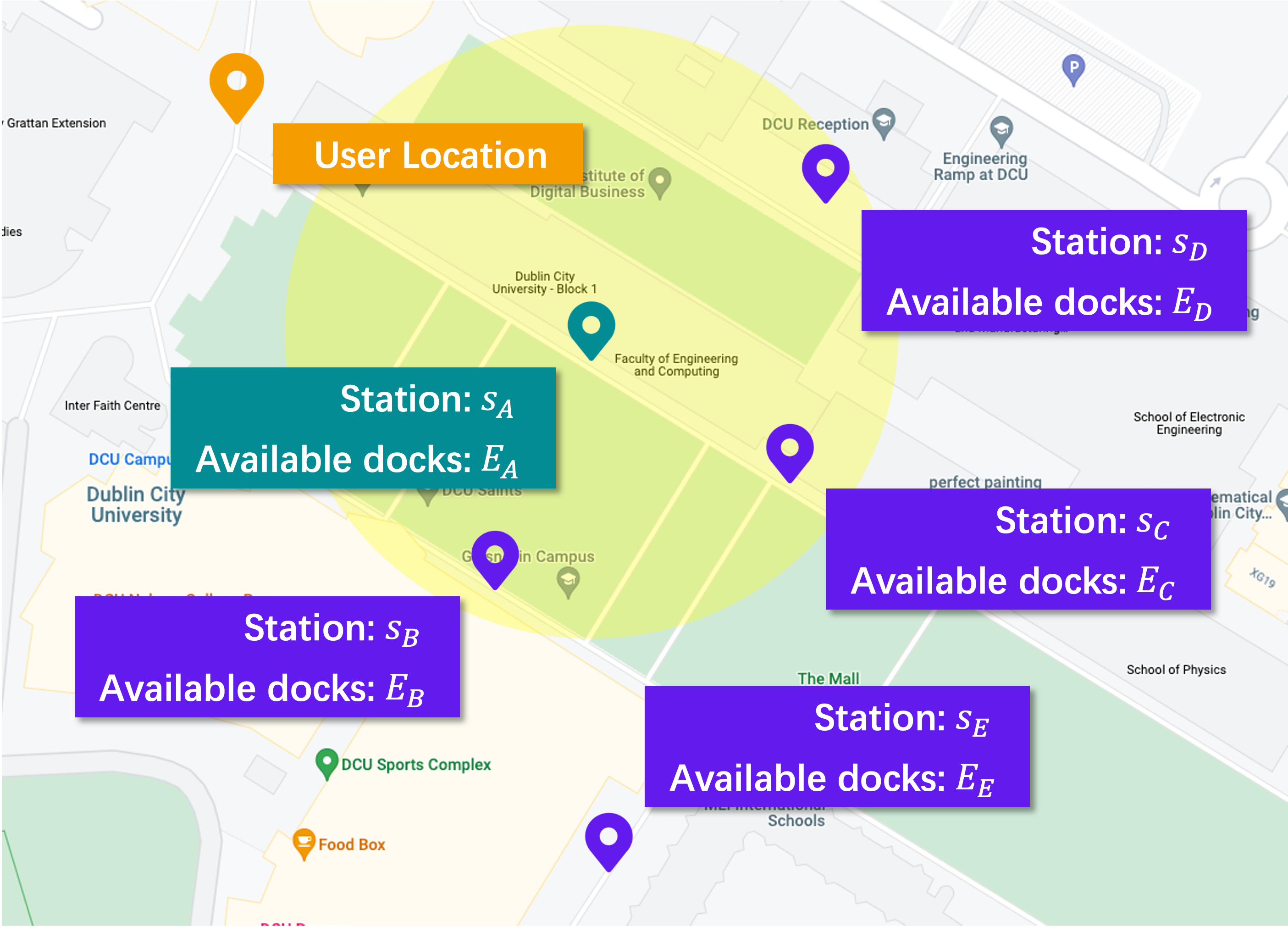}
            \caption{Parking station determination.}
            \label{fig: closest stations}
            \vspace{-0.1in}
        \end{figure}

        When a user confirms a destination predicted by U-Park or the \ac{ebike}'s location approaches the final predicted destination $\hat{d}$, within a distance of $dis$ metres, our final prediction module estimates parking space availability at nearby stations. This prediction is based on destination location, weather conditions, and estimated arrival time, as provided by the map \ac{api}. The results represent the availability of parking docks or spaces, helping to decide whether a particular station is feasible to recommend for parking. For instance, if a prediction suggests more than 5 available parking positions upon arrival, with an \ac{mae} of less than 1 \cite{Chen2021}, we are more confident in recommending that station, as it implies the user is likely to find suitable parking without difficulty. However, if the predictions indicate that there are fewer than, for example, 2 available parking positions, it implies that the user may have difficulty finding parking at their intended destination. In this case, U-Park will identify ``available stations" based on the predicted results of nearby stations. An ``available station" is defined as a station $s_u$ where the parking space availability $E_u$ exceeds a predefined threshold $E$. In \autoref{fig: closest stations}, we demonstrate the procedure for determining parking station $\hat{s_k}$ based on the probability distribution.
        
        In the scenario shown in \autoref{fig: closest stations}, Station $s_A$ represents the parking station at $\hat{d}$, with an associated available parking space denoted as $E_{\hat{d}}=E_A$. Stations $s_B$, $s_C$ and $s_D$ are the closest available stations to Station $s_A$, i.e., within the region of the yellow circle of radius $m$ metres. In contrast, Station $s_E$ lies outside the region and is not factored into the decision-making process. Formally, we express this as $\forall E_v>E$, where Station $s_v \in \{s_B, s_C, s_D\}$. In this scenario, we assume that $E_A<E$, indicating that Station $s_A$ is not available for parking. Thus, the parking station recommended to the user is chosen from only Stations $s_B$, $s_C$ and $s_D$. The probability of recommending Station $s_v$ to this user can be calculated by \eqref{equ: prob station}. For instance, the probability of Station $s_C$ being recommended to the user in this case is given by $P_C = E_C/(E_B+E_C+E_D)$.
        
        \begin{equation} \label{equ: prob station}
            P_v = \frac {E_v} {E_B+E_C+E_D} \qquad \text{$v \in \{B, C, D\}$}
        \end{equation}
        
        The parking station recommendation method based on the probability works better than simply advising every cyclist with the same destination to park at Station $s_v$ because it avoids the rapid change in parking availability at Station $s_v$ caused by the same recommendations. Instead, it helps balance the parking availability within a specific area.

\subsection{Trading Module \& Database} \label{subsec: arc-bc}
    At the beginning of a trip, the trading module charges the user for the trip deposit as per the prediction module result. At the end of the trip, users are billed for trip duration and parking behaviour. For instance, parking at suggested stations may earn users rewards. Data from the hardware, user interface, and trading modules is stored in a database, enabling interactions among these components.
    
    \section{System Implementation} 
        \label{sec: implement}
        In this section, we introduce and provide an overview of the implementation of the proposed system, U-Park, within the context of a shared \ac{ebike} system as a use case. 

\subsection{Hardware Module} \label{subsec: imp-pi}
    The hardware section is implemented using various sensors connected to the \ac{ebike} via a central controller, the \textit{Raspberry Pi 3 Model B}, which is capable of processing the physical status information collected by the sensors, including \ac{gps} time series and digital lock status of the \ac{ebike}. The \ac{gps} time series, required for our prediction module, is then transmitted to our database for further use, while the digital lock status is used for interactions between the mobile application and the Raspberry Pi to lock or unlock an \ac{ebike}. Specifically, the \ac{ebike}'s geographic location is identified by \textit{NEO-6M} GPS module, and its axial position is captured by the \textit{MPU6050} gyroscope and accelerometer module.

\subsection{User Interface Module} \label{subsec: imp-app}
\begin{figure}[htbp]
        \vspace{-0.1in}
        \centering
        \includegraphics[width=\linewidth]{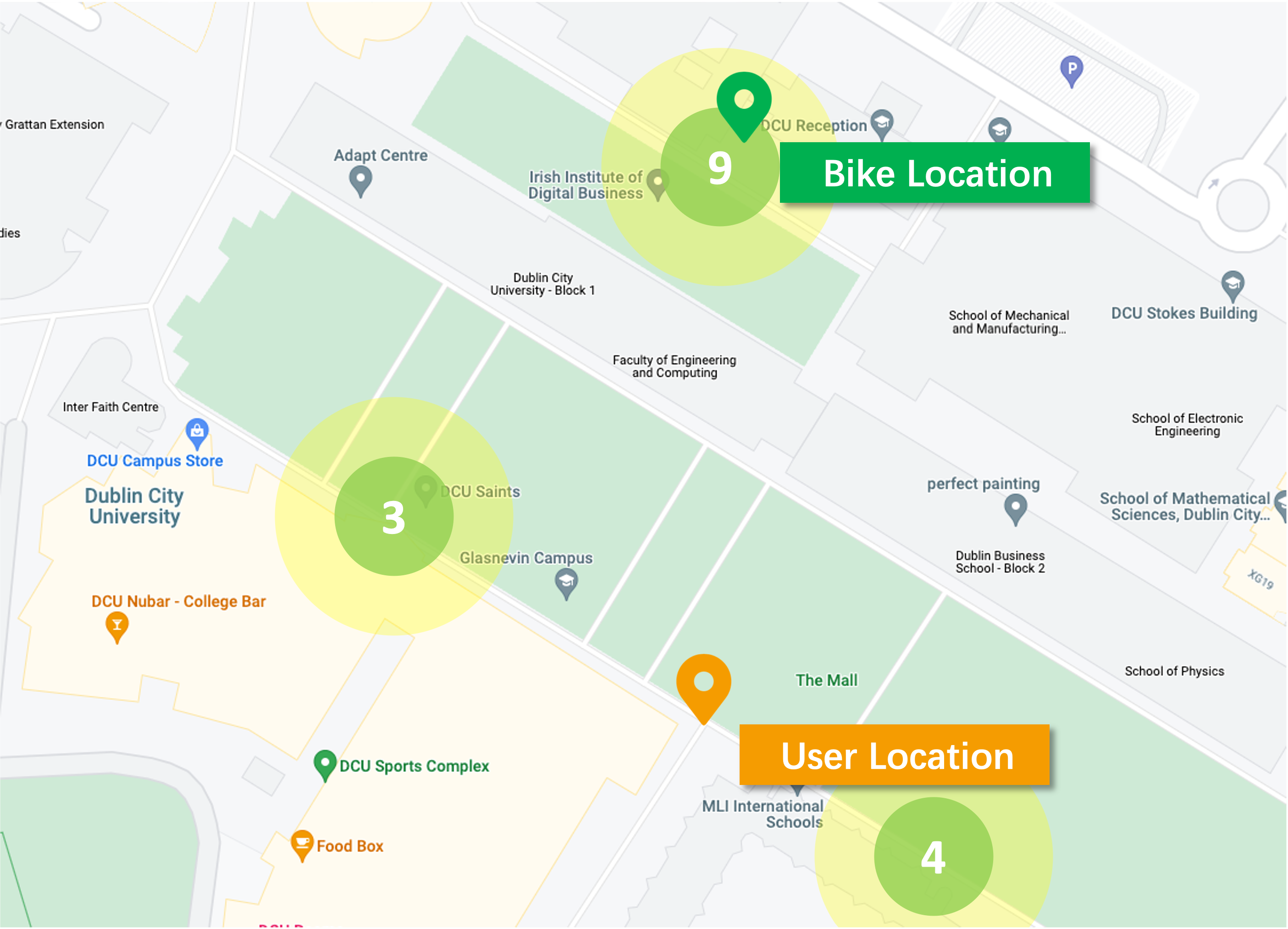}
        \caption{User location, \ac{ebike} position and predefined parking zones.}
        \label{fig: mobile-app-parking-zones}
        \vspace{-0.1in}
    \end{figure}

    Our Android mobile application is developed using \textit{IntelliJ IDEA} and installed and tested on an \textit{OnePlus 7pro} with a \textit{$Snapdragon^{TM}$ 855} processor. With Google Maps \ac{api}\footnote{\url{https://developers.google.com/maps/documentation/directions}}, the interactive map is implemented, presenting the locations of the user and \ac{ebike}s, as well as predefined station locations with circles representing their parking areas and digits showing parking availability in real-time. In \autoref{fig: mobile-app-parking-zones}, the orange marker represents the current user location, obtained via the Android locating service. The green marker shows the current location of an \ac{ebike}. Circular areas filled with different colours define parking zones with different parking fees, and the digit within each parking zone denotes its current parking availability. The size of parking zones is dynamically scaled to account for changes in \ac{gps} positioning accuracy in different areas for optimising performance in non-open areas.
    
    The \ac{qr} code scanning function, implemented by the \textit{ZXing} library\footnote{\url{https://github.com/dm77/barcodescanner}}, captures the \ac{ebike}'s \ac{uid}, \ac{gps} coordinates, and current timestamp. The \ac{ebike}'s \ac{uid} serves for management and maintenance, while the others are sent to our prediction module to predict the destination, which is subsequently recommended to the user for confirmation, and to examine whether the remaining battery power is enough for the predicted trip. If not, U-Park waits for the user's response and suggests an alternative \ac{ebike} nearby that has sufficient battery power. Conversely, if the battery power is enough, the predicted destination is used to calculate the trip deposit. The calculated deposit and the user's real-time credit are then transmitted to the database for validation. Data transmission between the user-end and the trading module is completed based on \textit{Web3j} \ac{api}\footnote{\url{https://docs.web3j.io/4.10.0/}}. 
    
    We used \textit{DynamoDB} for data storage and hardware testing. Interactions with \textit{DynamoDB} are facilitated through the \textit{Amplify} framework provided by \textit{Amazon Web Services}\footnote{\url{https://aws.amazon.com/dynamodb/}} for database management. The \ac{gps} data contains trip details, including trip ID, latitude, longitude, and timestamp. Additionally, we add extra information when a trip ends, including the moving duration, moving fee, minimum distance to stations, and parking fee. This enables us to validate trip deposits and balances. Trip records also include rental and return timestamps, along with \ac{gps} coordinates and physical attributes. A statistical description of this dataset is shown in \autoref{table: report collected}.
    
    \begin{table}
        \caption{Statistical description of test trip report collected.}
        \vspace{-0.1in}
        \label{table: report collected}
        \begin{tabularx}{\linewidth}{@{\extracolsep{\fill}}c c c c c}
            \toprule
            Object & t (s) & F$_m$ & d (m) & F$_p$ \\
            \midrule
            mean & 83.50 & 0.0025 & 36.16 & 0.0636 \\
            std & 89.32 & 0.0027 &   33.27 & 0.0414 \\
            min & 11.00 & 0.0003 & 1.94 & 0.0000 \\
            25\% & 24.00 & 0.0007 & 10.61 & 0.0500 \\
            50\% & 30.50 & 0.0009 & 20.35 & 0.0750 \\
            75\% & 112.50 & 0.0034 & 56.75 & 0.1000 \\
            max & 318.00 & 0.0095 & 95.19 & 0.1000\\
            \bottomrule
        \end{tabularx}
        \begin{tablenotes}\footnotesize
            \item[*] t (s) -- Trip Duration in Seconds, F$_m$ -- Moving Fee, d (m) -- Parking Distance in metres, F$_p$ -- Parking Fee.
        \end{tablenotes}
        \vspace{-0.2in}
    \end{table}

\subsection{Trading Module \& Database} \label{subsec: imp-bc}
    Since our paper focuses more on the design of the recommendation system, the trading module and database can be realised using various technologies depending on specific project requirements. Therefore, we will only discuss the management of user credits in this section.
    
    According to the incentive and punishment policy we introduced in \autoref{sec: review}, the parking fee $F_{parking}$ mentioned in \autoref{sec: design} and applied in U-Park is determined by a tiered fee approach, as shown in \eqref{equ: parking-fee}, where $D_1$ and $D_2$ are two distance thresholds and $D_{min}$ represents the minimum distance, measured in metres, between the \ac{ebike}'s current \ac{gps} location and the locations of all available stations. $F_{penalty}$ is the penalty imposed for the failure to park the \ac{ebike} appropriately.

    \begin{equation} \label{equ: parking-fee}
        F_{parking} =
        \begin{cases}
            0                       & \text{$D_{min} \leq D_1$} \\
            0.5 \times F_{penalty}  & \text{$D_1 < D_{min} \leq D_2$} \\
            F_{penalty}             & \text{$D_{min} > D_2$}
        \end{cases}
    \end{equation}
    
    According to \cite{Bohannon2011}, the lowest mean walking speed indicates that a 50-metre commute takes less than 1 minute, leading to a reward of \euro 0.2 in our study. Considering potential user heterogeneity, we set $D_1$ and $D_2$ to 50 and 100 metres, respectively. As an illustrative example, this choice results in corresponding penalties for user credits. To clarify, parking within the green circular zones in \autoref{fig: mobile-app-parking-zones} means that $D_{min} < 50$, leading to a parking penalty $F_{parking}$ of 0 for this trip. Parking within the yellow circular zones leads to half of the penalty while parking outside results in the full penalty charge.

\subsection{Dataset and Preprocessing} \label{subsec: imp-data}

    \subsubsection{Dataset 1: MOBY Dataset}
    
        We assessed our system's performance using the dataset obtained from MOBY Bike, a shared \ac{ebike} company\footnote{\url{https://mobybikes.com}}, as described in \cite{Yan2022}. Features used in our system include trip origin and destination labels, starting time, i.e., rental time, hourly rental demand at the starting station (derived from statistics), and weather conditions (e.g., precipitation amount and air temperature) collected from the \textit{Irish Meteorological Service}\footnote{\url{https://www.met.ie/climate/available-data/historical-data}}.
        
\begin{figure}[htbp]
            \vspace{-0.1in}
            \centering
            \includegraphics[width=\linewidth]{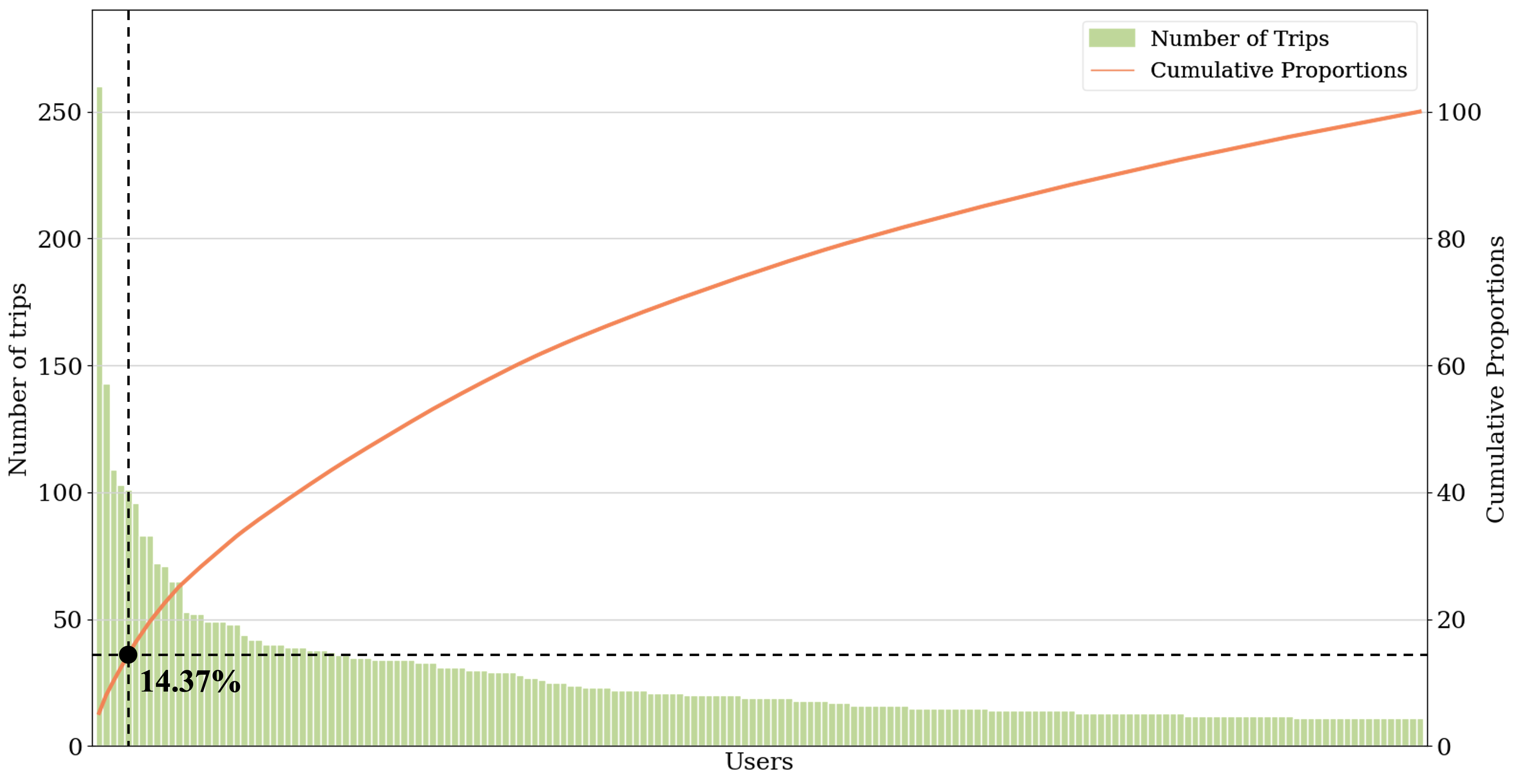}
            \caption{The number of trip records for each user.}
            \label{fig: user trips}
            \vspace{-0.1in}
        \end{figure}

        To prepare the dataset for model implementation, we considered all user data for the history-based model, filtered out users with less than 10 trip records in our dataset, and then selected the data from the top five users with the most trips for the trajectory-based prediction model. As shown in \autoref{fig: user trips}, these users contributed 716 complete journey records in total, occupying about 14.37\% of the total trips, so we believe that such experiments can reflect the performance of our proposed model to a certain extent. 
    
        \autoref{table: user info} illustrates the diverse trip patterns among the selected users. Specifically, Users 1, 3, and 5 have a similar number of origins and destinations within their trip histories. In contrast, there is a significant imbalance in the numbers of origins and destinations for Users 2 and 4. The difference in trip patterns could be a challenge for the prediction model. It also highlights the necessity of effective strategies that can accommodate and adjust individual user behaviours.
        
        \begin{table}[htbp]
            \caption{Statistics for selected user data.}
            \vspace{-0.1in}
            \label{table: user info}
            \begin{tabularx}{\linewidth}{@{\extracolsep{\fill}}c c c c c}
                \toprule
                user ID & trips & origins & destinations & proportions \\
                \midrule
                user 1 & 260 & 25 & 25 & 0.0522 \\
                user 2 & 143 & 2 & 11 & 0.0287 \\
                user 3 & 109 & 40 & 31 & 0.0219 \\
                user 4 & 103 & 18 & 9 & 0.0207 \\
                user 5 & 101 & 26 & 14 & 0.0203 \\
                \bottomrule
            \end{tabularx}
            \vspace{-0.1in}
        \end{table}
        
        To ensure compliance with \ac{gdpr} regulations, we performed the following steps: 1) splitting the feature \textbf{Starting Time} into separate components for the hour and weekday, and 2) transforming \ac{gps} information into road segment data, containing only road labels rather than exact \ac{gps} coordinates, by \textit{Mapbox Map Matching \ac{api}}\footnote{\url{https://docs.mapbox.com/api/navigation/map-matching}}. 
        
        It is worth mentioning that a significant imbalance in trip origins and destinations was observed for these users. To address this issue, we employed the \ac{smote} algorithm \cite{Chawla2002}. \ac{smote} works by calculating the $K$ nearest neighbours of each minority class sample, randomly selecting $N$ samples from these neighbours for random linear interpolation, and generating new minority class samples accordingly. This process resulted in an augmented training set with more samples in minority categories, addressing the imbalance problem. The modified dataset is presented in \autoref{table: modified dataset}, based on which \ac{ml} models were designed, implemented, and evaluated.

        \begin{table*}
            \caption{Sample data of trip records from the modified dataset.}
            \vspace{-0.1in}
            \label{table: modified dataset}
            \begin{tabularx}{\linewidth}{@{\extracolsep{\fill}}l c c c c c c c c c c c}
                \toprule
                O & D & W & H & Hourly rentals & PA & AT & WBT & DPT & VP & MSLP & Road segment\\
                \midrule
                215 & 199 & 1 & 16 & 2 & 0.0 & 21.0 & 18.4 & 16.7 & 19.1 & 1019.2 & \dots, Stillorgan Road, \dots\\
                215 & 214 & 5 & 14 & 2 & 0.0 &  5.9 &  4.7 &  3.1 &  7.6 & 1009.6 & \dots, Stillorgan Road, \dots\\
                261 & 261 & 0 & 20 & 2 & 0.0 & 13.5 &  7.9 &  0.1 &  6.2 & 1016.9 & \dots, Queen's Road, \dots\\
                217 & 261 & 5 & 19 & 1 & 0.0 &  9.5 &  7.5 &  5.1 &  8.8 & 1006.8 & \dots, Harbour Road, \dots\\
                261 & 210 & 2 & 13 & 1	& 0.0 & 18.4 & 14.8 & 12.0 & 13.9 & 1024.8 & \dots, Idrone Lane, \dots\\
                \bottomrule
            \end{tabularx}
            \begin{tablenotes}\footnotesize
                \item[*] O -- Rental ID, D -- Return ID, W -- Rental Weekday, H -- Rental Hour, PA -- Precipitation Amount (mm), AT -- Air Temperature (C), WBT -- Wet Bulb Temperature (C), DPT -- Dew Point Temperature (C), VP -- Vapour Pressure (hPa), MSLP -- Mean Sea Level Pressure (hPa).
            \end{tablenotes}
            \vspace{-0.2in}
        \end{table*}

    \subsubsection{Dataset 2: Dublinbikes Dataset}
        Inspired by the successful application of the \ac{astgcn} model in bike availability prediction, as detailed in \cite{Chen2021_1}, we initially intended to adapt its structure to predict parking space availability in our study. However, the MOBY dataset lacks specific data on parking space availability at the parking stations. To address this, we assumed that the distribution of available parking spaces in our dataset would resemble that of a comparable open dataset from NOW Dublinbikes\footnote{\url{https://www.dublinbikes.ie/en/home}}, used in \cite{Chen2021_1}. We then aligned the parking stations from both datasets based on geographical proximity.

\begin{figure}[htbp]
            \vspace{-0.1in}
            \centering
            \includegraphics[width=\linewidth]{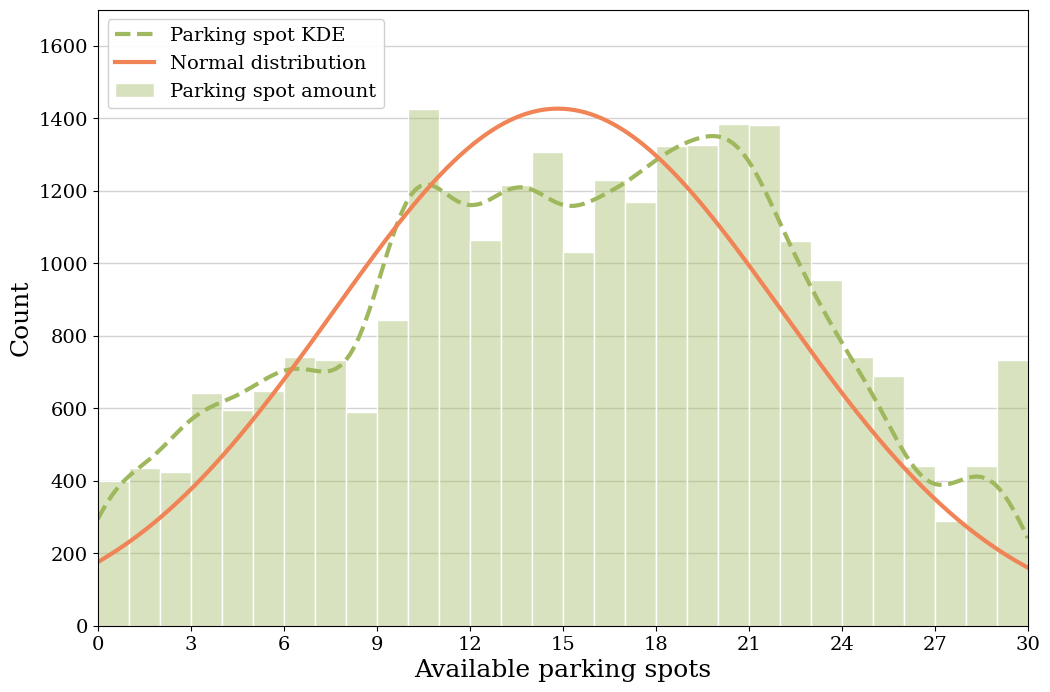}
            \vspace{-0.3in}
            \caption{Histogram of parking space in station 32 on York Street East.}
            \label{fig: user dist}
            \vspace{-0.2in}
        \end{figure}

        The statistical histogram of parking space is illustrated in \autoref{fig: user dist}. Accordingly, we approximate that the number of available spots follows a normal distribution, and then set up the baseline model for our study as follows: Initially, we model the actual number of parking spaces at each station using a normal distribution, i.e., $N \sim \mathcal{N}(\mu, \sigma^2)$, where $\mu$ is the average number of available spots and $\sigma$ represents its standard deviation. Next, we compute the probability that the simulated number of vacancies exceeds zero for each simulation run. Lastly, this process is repeated, ten thousand times in our case study, to derive the average probability of obtaining at least one available parking space.

\subsection{Prediction Module} \label{subsec: imp-model}
    As outlined in \autoref{sec: design}, our prediction module consists of three individual stages: destination prediction based on trip history, destination prediction based on partial trajectory, and parking space availability prediction based on station history. Each prediction task is addressed by various \ac{ml} models.
        
    \subsubsection{History-Based Destination Prediction} \label{subsubsec: imp-model-history}
        In our model, the dataset is divided into training, validation, and test sets with an 8:1:1 ratio, and a basic \ac{ann} model is implemented including two hidden layers with 64 and 32 units, respectively, and a dropout layer between them. The number of units in each layer is optimised by the Hyperband algorithm \cite{Li2017}, and they may vary when applied to other users' trip records. The loss function used is \ac{cce}, and the model performance is evaluated by Categorical Top-3 and Top-1 Accuracy metrics. To simulate the absence of certain features, we experimented with different combinations of feature sets in our work. The feature sets adopted are as follows: (1) rental location ID (Init); (2) rental hourly demand (D); (3) rental hour and rental weekday (T); (4) precipitation amount, air temperature, wet bulb temperature, dew point temperature, vapour pressure, and mean sea level pressure (W).
    
    \subsubsection{Trajectory-Based Destination Prediction} \label{subsubsec: imp-model-trajectory}
        Based on road segments transformed from the partial trajectory, we split the dataset into training data, validation data, and test data with an 8:1:1 ratio, and trained several models to predict the destination candidate. 

        \textit{Simple \ac{rnn}:} A simple \ac{rnn} model is implemented with a structure including a dense layer with 64 units, a dropout layer, and a dense layer with 32 units.

        \textit{A\ac{rnn}:} An \ac{arnn} model is implemented with a structure including a dense layer with 64 units, a Bahdanau attention layer, a dropout layer, and a dense layer with 32 units. 

        \textit{\ac{lstm}:} An \ac{lstm} model is implemented with a structure including an \ac{lstm} layer with 128 units, a dropout layer, and a dense layer with 32 units.

        \textit{A\ac{lstm}:} An \ac{alstm} model is implemented with a structure including an \ac{lstm} layer with 128 units, a Bahdanau attention layer, a dropout layer, and a dense layer with 32 units.
        
        The number of units in each layer is also tuned by the Hyperband algorithm. Specifically, for trips with a length of $L_k > (50+10-1)$, we randomly select 50 numbers smaller than $(L_k-4)$ and accordingly extracted 50 samples starting from the trip origin. Each sample is processed by considering only the first 4 and the last 4 road segments to create a sub-trajectory for this trip. For instance, for a trip $\mathbb{T}_k$ of which the length is 70, if a generated random number is 20, we take the first 20 points (i.e., $[(t_{1}^k,p_{1}^k),\dots,(t_{20}^k,p_{20}^k)]$) as one sample, and only use the first 4 (i.e., $[(t_{1}^k,p_{1}^k),\dots,(t_{4}^k,p_{4}^k)]$) and last 4 points (i.e., $[(t_{17}^k,p_{17}^k),\dots,(t_{20}^k,p_{20}^k)]$) to produce a sub-trajectory. For the selected user, 46 trips were suitable for this prediction task, resulting in 2,300 sub-trajectories containing the first four and last four road segment labels. We tested different feature combinations for this model. The initial feature combination (Init) is only a serial of road segment labels. Subsequently, we added temporal attributes (T) and weather conditions (W) as features. The loss function used is \ac{cce}, and the model's performance is assessed using Categorical Accuracy.
        
    \subsubsection{Parking Spot Prediction \& Recommendation} \label{subsubsec: imp-model-demand}
        Following the same experiment setup as in \ac{astgcn} \cite{Chen2021_1}, we use the data on a similar of available spots at each \ac{ebike} station over the initial 3-hour period to predict the parking space availability 45 minutes later. Specifically, each data point represents the average number of available spots calculated over 15-minute intervals, so the historical data length is $3 \times 60 \div 15 = 12$, while the prediction length is $45 \div 15 = 3$. The dataset is divided into training, validation, and test sets with a 6:2:2 ratio. Features used include the number of available spots, time of day, weekday, and weather condition description. The model is optimised by the Adam algorithm, and other parameters, such as learning rate, are set the same as introduced in \cite{Chen2021_1}.

        After obtaining the prediction results, we optimise our recommendation based on the strategy introduced in \autoref{sec: design} which considers the parking availability within a specific area rather than merely one parking station. The distance threshold $dis$ used in our experiments is 300, which means U-Park will find available stations as recommendation candidates within 300 metres of the destination and provide parking suggestions based on predicted parking spots.
    
    \section{Results \& Discussion} 
        \label{sec: result}
        \subsection{Feature Selection} \label{subsec: res-feature}

    We introduced various feature combinations in the previous section. To reduce the influence of random factors, we conducted 10 evaluations for each combination and recorded the mean values of each metric in \autoref{fig: ablation history} and \autoref{fig: ablation trajectory}. We present the results for five users for case-specific analysis. However, the outcomes may vary when considering other users.
    
    \subsubsection{History-based Prediction Model}
\begin{figure}[htbp]
        \vspace{-0.1in}
        \centering
        \includegraphics[width=\linewidth]{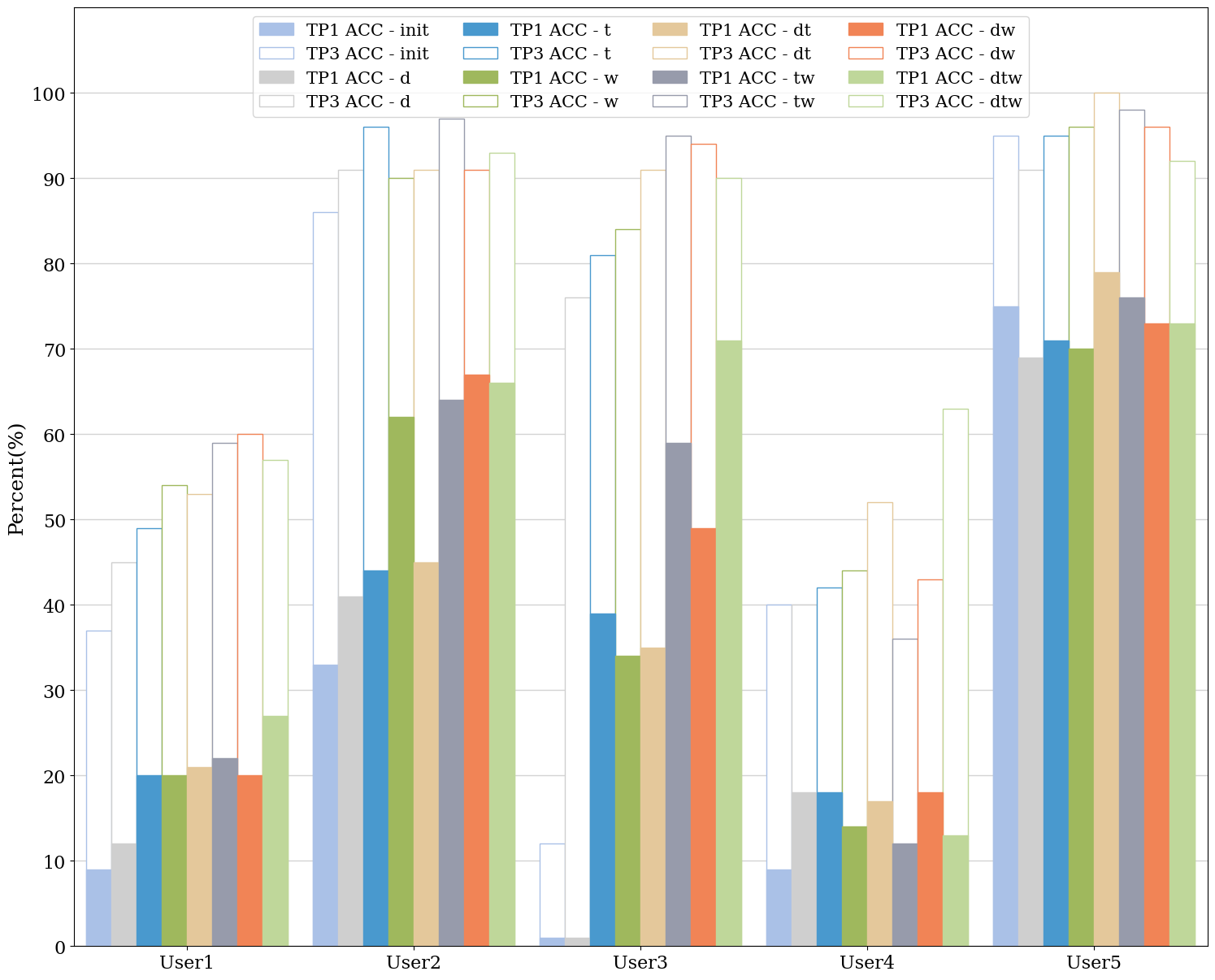}
        \vspace{-0.3in}
        \caption{History-based results with different features.}
        \label{fig: ablation history}
        \vspace{-0.1in}
    \end{figure}

        It is evident from \autoref{fig: ablation history} that the Top-3 accuracy consistently outperforms the Top-1 accuracy across all users and feature sets. Moreover, we found that including additional features beyond the initial rental location ID significantly improved model performance. For instance, for User 2, Top-3 and Top-1 accuracy increased by approximately 11\% and 31\%, respectively, when incorporating demand (d) and weather-related (w) features. However, the optimal feature set can vary depending on the user. Specifically, User 4 achieved the highest Top-1 accuracy (around 67\%) with a combination of rental location ID (init), hourly demand (d), and weather-related features (w).

    \subsubsection{Trajectory-based Prediction Model}
\begin{figure}[htbp]
        \vspace{-0.1in}
        \centering
        \includegraphics[width=\linewidth]{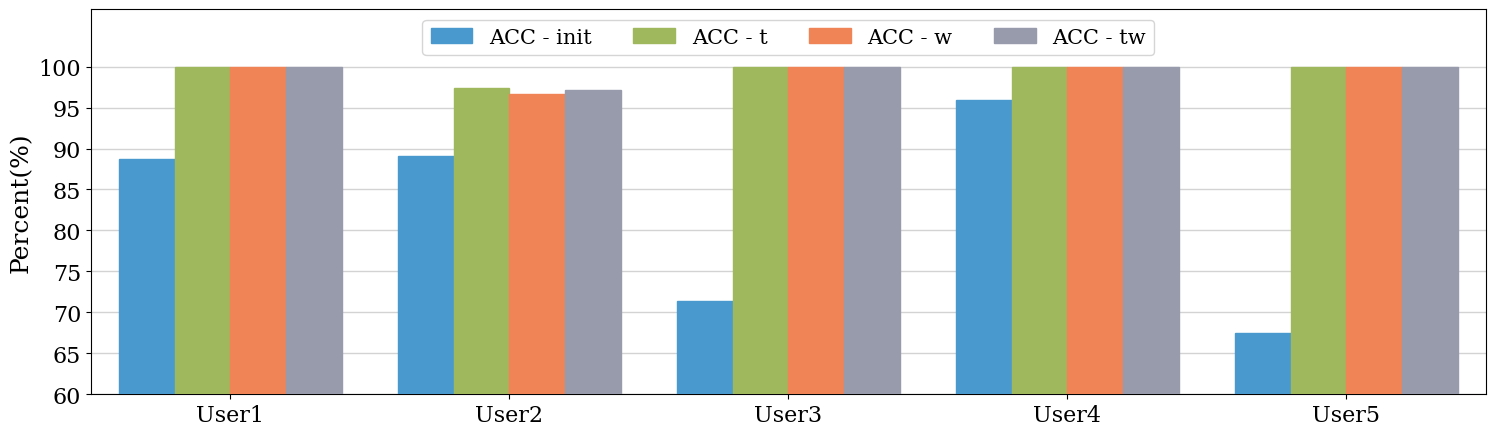}
        \vspace{-0.3in}
        \caption{Trajectory-based results with different features.}
        \label{fig: ablation trajectory}
        \vspace{-0.1in}
    \end{figure}

        \autoref{fig: ablation trajectory} also demonstrates that the performance of the trajectory-based model varies with different feature sets. Different to the history-based model, a combination of the initial input (road segment labels), time (t), and weather-relevant (w) attributes consistently achieved the highest accuracy across all selected users. Moreover, the trajectory-based model significantly outperforms the history-based model, reaching 100\% accuracy for the majority of users and achieving 97.60\% even for the relatively challenging case of User 2.

\subsection{Model Comparison for Trajectory-based Prediction} \label{subsec: res-model}
\begin{figure}[htbp]
        \vspace{-0.1in}
        \centering
        \includegraphics[width=\linewidth]{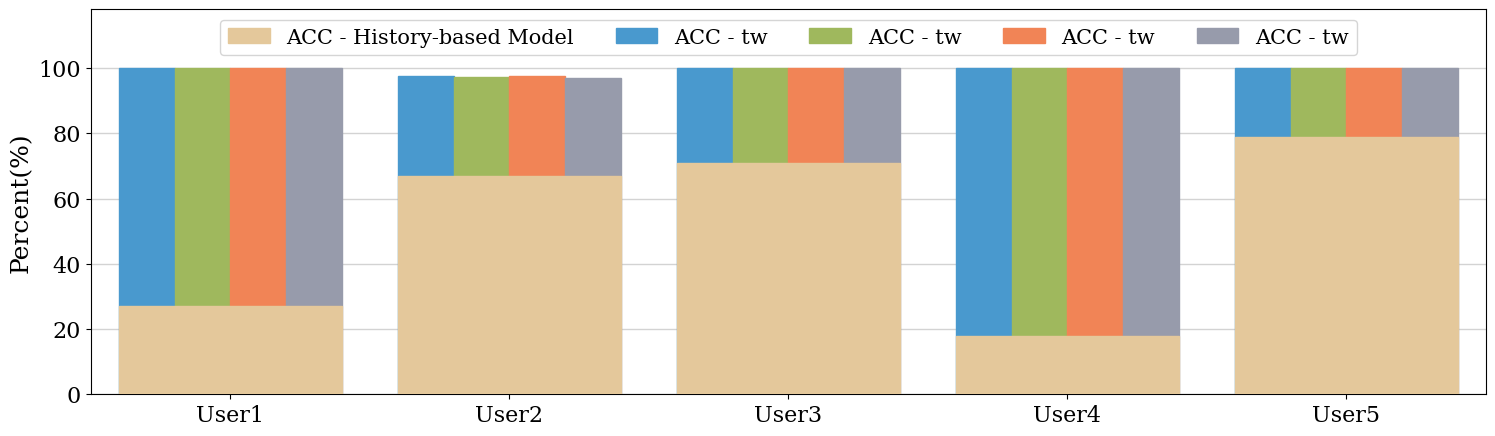}
        \vspace{-0.3in}
        \caption{Trajectory-based results with different models.}
        \label{fig: trajectory models}
        \vspace{-0.1in}
    \end{figure}

    \autoref{fig: trajectory models} Compares the performance of various trajectory-based models, including Simple \ac{rnn}, \ac{arnn}, \ac{lstm}, \ac{alstm}). It is obvious that incorporating trajectory information can significantly improve system performance, with accuracy improvements ranging from 29\% (User 1) to 82\% (User 4) compared to the model based on history. Interestingly, different \ac{ml} models exhibit varying effectiveness among different users. While all models achieve 100\% accuracy for certain users with consistent travel patterns, the best model may vary based on individual user behaviour.

\subsection{Selection of K-value}
    As mentioned above, the first $k$ and last $k$ points in the journey play a significant role in the prediction results. Thus, we conducted experiments to assess the impact of different values of $k$, i.e., input size, and summarised the results in Table \autoref{table: k value}, from which we found that the highest prediction accuracy was achieved when $k = 4$. Consequently, the optimal choice for k in this prediction task is 4, which balances the prediction accuracy and input size.

    \begin{table*}[htbp]
        \caption{Impact of input size on prediction results.}
        \vspace{-0.1in}
        \label{table: k value}
        \begin{tabularx}{\linewidth}{@{\extracolsep{\fill}}c c c c c c c c c c c}
            \toprule
            K Value & 1 & 2 & 3 & 4 & 5 & 6 & 7 & 8 & 9 & 10 \\
            \midrule
            Test Loss & 0.0856 & 0.1027 & 0.1113 & \textbf{0.0654} & 0.0828 & 0.0804 & 0.0694 & 0.0977 & 0.1069 & 0.1057 \\
            Test Accuracy & 0.9698 & 0.9609 & 0.9478 & \textbf{0.9789} & 0.9667 & 0.9774 & 0.9763 & 0.9641 & 0.9618 & 0.9571 \\
            \bottomrule
        \end{tabularx}
        \vspace{-0.2in}
    \end{table*}

\subsection{Parking Space Availability Prediction \& Impact}    
    
We illustrate the performance of our parking space availability prediction with examples from five parking stations in \autoref{table: ps performance}, which includes each station's location (latitude and longitude), basic statistical information of available spots (average value and standard deviations), as well as different \ac{mae} and \ac{rmse} metrics based on the predictor used.
    
    \begin{table}[htbp]
        \caption{Performance (\ac{mae} \& \ac{rmse}) of parking spot prediction.}
        \vspace{-0.1in}
        \label{table: ps performance}
        \begin{tabularx}{\linewidth}{@{\extracolsep{\fill}}c c c c c c c}
            \toprule
            \multirow{2}{*}{\textbf{ID}} & \multirow{2}{*}{\textbf{Latitude}}  & \multirow{2}{*}{\textbf{Longitude}} & \multicolumn{2}{c}{\textbf{Available Spots}} & \multicolumn{2}{c}{\textbf{Prediction Results}}\\
            \cmidrule(lr){4-5} \cmidrule(lr){6-7}
            & & & mean & std & \ac{mae} & \ac{rmse}\\
            \midrule
            1 & 53.3569 & -6.2680 & 14.4737 & 4.7863 & 0.8880 & 1.3993 \\
            2 & 53.3511 & -6.2700 & 14.4107 & 4.6208 & 1.0018 & 1.5732 \\
            5 & 53.3431 & -6.2707 & 14.3387 & 5.6082 & 0.8363 & 1.3962 \\
            6 & 53.3430 & -6.2733 & 23.7705 & 5.0778 & 0.8853 & 1.7610 \\
            8 & 53.3420 & -6.2648 & 13.2831 & 8.3851 & 2.0066 & 3.2325 \\
            \bottomrule
        \end{tabularx}
        \vspace{-0.2in}
    \end{table}

    \begin{figure}[htbp]
        \vspace{-0.1in}
        \centering
        \includegraphics[width=\linewidth]{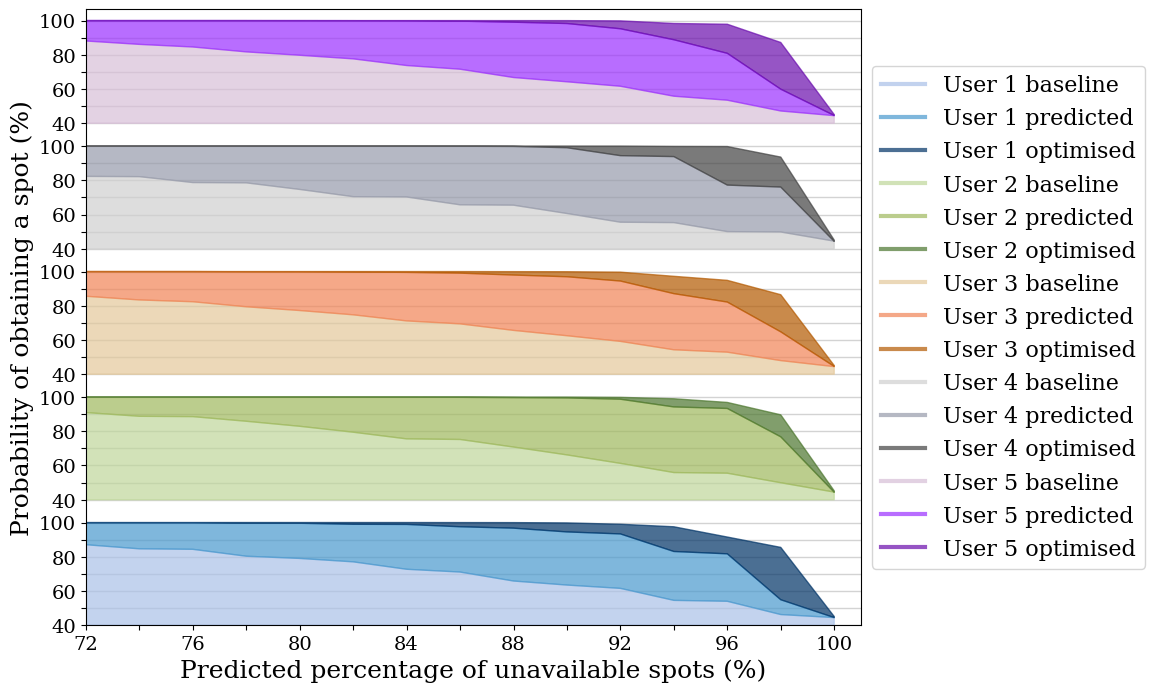}
        \caption{Impact of parking space prediction and optimisation.}
        \label{fig: system performance}
        \vspace{-0.1in}
    \end{figure}

    To best illustrate the efficacy of the proposed system, we evaluated the probability of finding a parking spot for five real users across three different settings: (a) without any recommendation system as a baseline (\textit{User X baseline}); (b) using the parking availability predictor (\textit{User X predicted}); and (c) using our U-Park system (\textit{User X optimised}), which recommends the station with the highest likelihood of finding a spot near the user's destination. Our key findings are shown in \autoref{fig: system performance}. Specifically, for a given predicted percentage of unavailable spots, we calculate the probability of finding a parking spot in each of the three settings for all five users. The probabilities vary by user, as they are averaged based on historical data from all stations each user has visited. Furthermore, our results demonstrate that the prediction module can significantly help improve the probability of all users obtaining a parking spot compared to the baseline result, thanks to its high prediction accuracy, i.e., low \ac{mae}. U-Park further optimises performance over the predicted results, especially when parking spot unavailability ranges from 88\% to 96\%, leading to an average improvement of 24.91\% and a peak of 29.66\% across the five real users analysed. Clearly, this demonstrates the efficacy of our system in real-world scenarios, especially where parking availability is limited.
    
    \section{Limitations} 
        \label{sec: limitation}
        Our framework proposed in this paper represents the initial stage of developing smart parking management for \ac{esms}. Although our system has been tested using real-world data, it has not been implemented in a practical context. The final prediction stage, focusing on parking space availability, relies heavily on real-world data from shared \ac{ebike} stations, which we had to approximate with comparable datasets during the study as an alternative. Reliable and comprehensive data are crucial for accurate predictions, but acquiring open data sources with detailed \ac{gps} time series and parking space information at each station remains a challenge. Incorporating an automated mechanism for model selection and hyperparameter optimisation would enhance our system. Our current work has primarily utilised simple \ac{ml} models due to the limited dataset size, as more complex models could lead to overfitting. Access to larger and more diverse datasets, along with advanced \ac{ml} models, is expected to substantially enhance the performance and universality of our recommendation system within the existing framework. Besides, given that most \ac{ebike} docking sites are generally located in open areas in our case study in Dublin, Ireland, our analysis was based on the assumption that it applies to these environments. Therefore, we did not specifically address non-open areas in our system.
    
    \section{Conclusion \& Future Work}
        \label{sec: conclusion}
        In this paper, we introduce U-Park, a smart parking management solution for \ac{esms}. We outlined its architecture, including two prediction tasks across three stages, to provide personalised parking recommendations to assist users. For the prediction of trip destinations in the first two stages, we significantly improved accuracy to over 97.60\% by employing a partial trajectory-based \ac{arnn} model, which dynamically refines predictions from a trip history-based model. Additionally, our parking space availability prediction model empowers U-Park to suggest the optimal parking station to users based on the availability of predicted parking spaces, improving the probability of obtaining a parking spot by 24.91\% on average and 29.66\% on maximum when parking availability is limited. 

In the future, our system can enhance its performance by integrating more advanced \ac{ml} algorithms and well-structured models. We also intend to evaluate our system using larger and more suitable datasets and potentially deploy it in real-world industry scenarios to validate its effectiveness. 
    
    \section*{Acknowledgement} 
        \label{sec: acknowledgement}
        The authors extend their gratitude to the Smart DCU Programme and the technical support team from MOBY Bikes for providing the data and engaging in valuable discussions. Additionally, they appreciate the assistance from Rakesh D. Murthy for his contributions to the trading module and database implementation. This research was conducted with the financial support of Science Foundation Ireland \textit{21/FFP-P/10266} and \textit{12/RC/2289\_P2} at Insight the SFI Research Centre for Data Analytics at Dublin City University. The data used in this work is fully anonymous. For the purpose of Open Access, the author has applied a CC BY public copyright licence to any Author Accepted Manuscript version arising from this submission.

    \bibliographystyle{IEEEtran}
    \bibliography{reference}
    
    %\section*{\textcolor{magenta}{Biography Section}}
       % \label{sec: biography}
        %\input{biography}

\end{document}